\documentclass[galaxies,aricle,accept,pdftex,moreauthors]{Definitions/mdpi}

\firstpage{1} 
\pubvolume{1}
\issuenum{1}
\articlenumber{0}
\pubyear{2022}
\copyrightyear{2022}
\externaleditor{Academic Editor: {Mubasher Jamil and Tao Zhu} 
} 
\datereceived{14 June 2022} 
\dateaccepted{22 August 2022} 
\datepublished{} 
\hreflink{https://doi.org/} 

\Title{Properties of MAXI J1348-630 during Its Second Outburst in~2019}
\TitleCitation{Properties of MAXI J1348-630 during Its Second Outburst in 2019}

\Author{Riya Bhowmick 
 $^{1}$\orcidA{}, Dipak Debnath $^{2,1,}$  
 *\orcidB{}, Kaushik Chatterjee $^{1,2}$\orcidC{}, Arghajit Jana $^{3}$\orcidD{} and Sujoy Kumar Nath $^{1}$\orcidE{}}

\AuthorNames{Riya Bhowmick, Dipak Debnath, Kaushik Chatterjee, Arghajit Jana and Sujoy K. Nath}
\AuthorCitation{Bhowmick, R.; Debnath, D.; Chatterjee, K.; Jana, A.; Nath, S. K.} 

\address{%
$^{1}$ \quad Indian Centre for Space Physics, 43 Chalantika, Garia St. Rd., Kolkata 700084, India; riyabhowmickmalda@gmail.com~(R.B.); mails.kc.physics@gmail.com~(K.C.); sujoynath0007@gmail.com~(S.K.N.)\\
$^{2}$ \quad Institute of Astronomy Space and Earth Science, AJ 316, Sector II, Salt Lake, Kolkata 700091, India\\
$^{3}$ \quad Institute of Astronomy, National Tsing Hua University, Hsinchu 30013, Taiwan; argha0004@gmail.com~(A.J.)}

\corres{Correspondence: dipakcsp@gmail.com}

\abstract{The newly discovered galactic black hole candidate (BHC) MAXI~J1348-630 showed two major outbursts in 2019, just after its discovery.
Here, we provide a detailed spectral and temporal analysis of the less-studied second outburst using archive data from multiple satellites, 
namely Swift, MAXI, NICER, NuSTAR and AstroSat. The outburst continued for around two and a half months. Unlike the first outburst from this source, 
this second outburst was a `failed' one. The source did not transition to soft or intermediate spectral states. During the entire outburst,
the source was in the hard state with high dominance of non-thermal photons. The presence of strong shocks are inferred from spectral
fitting using a TCAF model. In NuSTAR spectra, weak reflection is observed from spectral fitting. Low-frequency quasi-periodic
oscillations are also detected in AstroSat data.}


\begin{document}
\section{Introduction}
\label{sect:intro}

Black hole X-ray binaries are one of the most interesting astronomical objects as they show rapid variability in timing and spectral properties. 
A black hole X-ray binary system consists of a black hole (BH) and a companion star. At~some later stage during evolution, the~companion star fills 
its Roche lobe and, due to the immense gravitational pull of the BH, mass from the companion starts to accrete towards the black hole through the 
Lagrangian point L1~\citep{Frank02}. This process is known as the Roche lobe overflow. There may also be wind accretion and tidal deformation.
Matter accretes to the BH after forming a spirally rotating accretion disk around the BH. BHs can be explored by detecting the electromagnetic
radiation coming out from the accretion disk around them. The~gravitational potential energy of the accreting matter is converted into energy that is
radiated over the entire electromagnetic wave band (from radio to $\gamma$-ray). 

Stellar mass black hole X-ray binaries can be classified as either transient or persistent.  
Persistent sources always remain in the active 
phase, whereas transient sources mostly remain in the quiescent phase and occasionally have outbursts.  
In~the case of transient sources,
viscosity plays a major role in triggering an outburst. The~inflowing matter from the companion initially accumulates at the `pile-up' radius~\citep {SKC19, RB21, KC22}, and when the viscosity increases and exceeds the critical limit, the~accumulated matter starts to accrete and triggers an outburst. The~outbursts of X-ray novae can be well-explained by the hydrogen instability 
model~\citep{Cannizzo1982, Dubus2001, Lasota2001}. In~the case of low-mass X-ray binaries, the~hydrogen ionization instability operating
in an accretion disk can describe the shape of outburst light curves~\citep{Janiuk2011, Baginska2021}.

An outbursting BH generally goes through four spectral states: low hard state (LHS) or hard state (HS), hard intermediate state (HIMS), soft intermediate 
state (SIMS) and soft state (SS) or high soft state (HSS)~\citep{RM06,MR09,DD13}. Evolution of states can be observed through a hardness intensity diagram (HID)  or ``q'' diagram~\citep{Belloni05,Belloni10} and an accretion-rate ratio intensity diagram~\citep{AJ16,KC20}. A~BH is observed with low luminosity in the
HS at the beginning of an outburst. As~time passes, 
luminosity increases and the BH moves towards the SS via HIMS and SIMS. In~SS, the luminosity remains very high. After~that, the~source enters the 
declining phase and moves to the HS through SIMS and HIMS. In~short, the~BH evolves through the four canonical spectral states, forming a hysteresis loop 
in the sequence: HS (rising) $\rightarrow$ HIMS (rising) $\rightarrow$ SIMS (rising) $\rightarrow$ SS $\rightarrow$ SIMS (declining) $\rightarrow$ 
HIMS (declining) $\rightarrow$ HS (declining).
An outburst can be classified as normal or failed, depending on its spectral variation~\citep{DD17}. Normal outbursts evolve
through all the spectral states and are complete in nature. Failed outbursts do not show softer spectral states (SIMS and SS). 
 {Normal outbursts are sometimes defined as `successsful' outbursts, and failed outbursts are 'hard-only' outbursts}~\citep{Tetarenko16}.

Generally, the energy spectrum of a BH consists of two components: a multi-color thermal black body or disk black body (DBB) and a non-thermal 
powerlaw (PL). Non-thermal, high-energy radiation dominates the harder states (HS and HIMS), and~thermal black body radiation dominates 
the softer states (SIMS and SS). Multi-color black body emission originates in the standard disk~\citep{NT73, SS73}, 
and powerlaw emission originates in a Compton cloud consisting of hot electrons~\citep{ST80, ST85}. Various models are present 
in the literature to understand and explain the accretion properties of black holes, such as Bondi flow~\citep{Bondi1952}, standard disk model 
\citep{SS73}, thick disk model~\citep{Paczynski80} and ADAF model~\citep{Narayan94}. These models can explain the radiation 
spectra of black holes to some extent. In~the mid 1990s, Chakrabarti and his collaborators came up with the Two Component Advective Flow (TCAF) 
solution based on transonic flow and radiative transport equations~\citep{CT95, SKC97, SKC16}. In~this model, 
the accretion flow consists of two components: a geometrically thin, optically thick, high-viscosity Keplerian disk, and a low-viscosity, optically thin, 
sub-Keplerian flow or halo. The~Keplerian matter accretes on the equatorial plane and is immersed within the sub-Keplerian flow. The~sub-Keplerian
flow (halo) moves in freefall timescale, and it moves faster than the Keplerian matter moving in a viscous timescale. The~sub-Keplerian flow
temporarily slows down at the centrifugal barrier and forms an axisymmetric shock~\citep{SKC90}. 
The post-shock region is hot and puffed-up and is known as the CENtrifugal pressure supported BOundary Layer (CENBOL). The~CENBOL acts as a Compton 
cloud in the TCAF solution. Multi-color black body spectra are generated from the soft photons originating in the Keplerian disk. A~fraction of
these soft photons (from the Keplerian disk) are intercepted by the CENBOL and are inverse-Comptonized by highly energetic `hot' electrons of 
the CENBOL to produce hard photons. The~powerlaw tail in the spectra is produced by these hard photons. Part of the hard photons interact
with the Keplerian disk, and for this reason a 'reflection hump' is observed at high~energy. 

The TCAF model has been implemented as an additive table model in XSPEC to obtain direct estimation of flow parameters and mass of the BH from spectral 
fitting~\citep{DD14}.  
This model has four basic flow parameters: (i) Keplerian disk rate ($\dot{m}_d$ in $\dot{M}_{Edd}$), (ii) sub-Keplerian
halo rate ($\dot{m}_h$ in $\dot{M}_{Edd}$), (iii) shock location ($X_s$ in Schwarzschild radius $r_s$), i.e.,~the boundary of CENBOL, and (iv) compression 
ratio (R), i.e.,~the ratio of the pre-shock matter density to the post-shock matter density and two other parameters: mass of the BH ($M_{BH}$ in 
$M_{\odot}$) and normalization (N). From~recent studies by our group, we can claim that the TCAF model is quite successful for explaining the physics 
around compact objects. Accretion flow dynamics of black hole candidates (BHCs) can be understood more clearly from the analysis of more 
than fifteen BHCs~\citep{DD14, DD15a, DD15b, DC16, DC19, DC21, KC20, KC21, AJ16} using the TCAF model. The~mass of BHCs 
has also been estimated from spectral analysis using the TCAF model~\citep{Molla16, Molla17, AJ20, SN22}. 
  
In BHs, jets/outflows are very important phenomena. In~astrophysical jets, mass, energy and angular momentum are channeled as a beam of ionized matter 
along the axis of rotation. Jets from compact objects are geometrically narrow and conical in shape. In~HS, collimated and compact jets are observed, whereas in intermediate states (HIMS and SIMS), jets are observed as discrete and blobby in nature. The~contribution of the jet component to the total observed 
X-ray flux can be estimated from spectral analysis with the TCAF model~\citep{AJ17, DD21}. Generally, no jets are observed in SS. 
However, there can also be jets in SS for magnetically dominated accretion disks~\citep{DD21}. {The precise mechanism of the
production of jets is still unknown. Though~jets are widely observed, the reasons behind their creation, collimation and~acceleration are still up for debate. Several theories have been proposed in the literature to explain jets and outflows, such as de-Laval nozzles}~\citep{Blandford1974},
{an electrodynamical acceleration model}~\citep{Znajek1978} and {self-similar centrifugally driven outflows}~\citep{Blandford1982}. Blandford and Znajek 1977~\citep{Blandford1977}
{made one of the earliest models for jet solutions. The~Blandford--Znajek process describes how jet power is extracted from the spin energy of the black
hole. A jet will be visible in all spectral states if it is driven by spin energy. However, in reality, jets are not observed in the soft
spectral state. Generally, it is thought that the magnetic field causes jets to collide}~\citep{Camenzind1989}. According to Chakrabarti and Bhaskaran 1992~\citep{SKC92}, {jets or outflows are ejected, accelerated and collimated by hydromagnetic processes.} In the TCAF model, the~CENBOL acts as the base of the jets. Here, radiation pressure is responsible for launching the jet~\citep{SKC99}. 

BHs exhibit quasi-periodic oscillations (QPOs) in some spectral states. {Low-frequency QPOs (LFQPOs) are very common observable features in the power density spectrum (PDS) of stellar mass BHs. A~few BHs show high-frequency QPOs in their PDSs. These X-ray transient sources exhibit QPOs with frequencies ranging from mHz to a few hundred~Hz}~\citep{Morgan1997}. {Many scientific groups have reported low- ($\sim$0.01--30~Hz) as well as high- \mbox{($\sim$40--450~Hz)} frequency QPOs in black hole X-ray binaries (BHXRBs)} (for a review
see~\citep{RM06, DD13} and references therein). Depending on their 
nature (Q value, RMS amplitude, noise, etc.), LFQPOs can be divided into three types: type-A, type-B and type-C~\citep{Casella2005, Motta2016}. In~HS and HIMS, 
type-C QPOs can be seen. In~SIMS, type-A or type-B QPOs are observed. Generally QPOs are not observed in SS. The~origin of the QPOs can be described as the oscillation of the CENBOL in the TCAF model~\citep{DD14, Mondal15, SKC15, DC16}. 
When the radiative cooling timescale and the infall timescale roughly match, the~outer boundary of the CENBOL oscillates, and~the emerging photons 
produce QPOs~\citep{MSC96, Ryu97, SKC15}. QPOs are also observed if Rankine--Hugoniot conditions for the 
stable shock are not satisfied~\citep{Ryu97}.   

The Galactic BHC MAXI~J1348-630 was discovered on 26 January 2019 by the gas slit camera (GSC) onboard Monitor of All-sky X-ray Image (MAXI)
\citep{Yatabe2019, Matsuoka09}, and based on its estimated mass and spectral features, it was classified as a BH binary~\citep{Tominaga2020}.
Swift/XRT observation localized the source at R.A.~=~$13^h$48$'$12.73$''$, Decl.~=~$-63^{\circ}$16$'$26.8$''$~\citep{Kennea2019}
 The~source was also observed
by several X-ray observatories, such as INTEGRAL, NICER, Insight-HXMT, AstroSat, Swift and NuSTAR~\citep{Lepingwell19, Sanna19, Chen19}. 
After completion of the first outburst, there was a quiescence of $\sim$20 days, and the source again re-brightened and started a new outburst that lasted for two and a half months from MJD$\sim$58,630 to MJD$\sim$58,700. After~completion of this second large outburst, six mini outbursts were
detected during the centroid time of MJD$\sim$58,747.6~$\pm$~0.4, MJD$\sim$58,812.7~$\pm$~0.3, MJD$\sim$58,886.8~$\pm$~0.4, MJD$\sim$58,975.8~$\pm$~1.0,
MJD$\sim$59,033.3~$\pm$~1.6 and MJD 59,098.0~\citep{Negoro2020, Baglio2020, Carotenuto2020}.

The first outburst lasted for four months, and the source showed all four canonical spectral states. A~detailed study of the spectral
evolution of the first outburst was performed with MAXI~\citep{Tominaga2020, AJ20}, Swift~\citep{AJ20, Zhang2022} and Insight-HXMT
\citep{Zhang2022}. Chauhan~et~al. (2020)~\citep{Chauhan2020} estimated the most-probable distance of MAXI J1348-630 as $2.2^{+0.5}_{-0.6}$ kpc. Based on spectral analysis,
Tominaga~et~al. (2020)~\citep{Tominaga2020} predicted that MAXI J1348-630 hosts a relatively massive black hole and reported a range of mass of the BH dependent on different 
spinning parameters and inclination angles. However, Jana~et~al., 2020~\citep{AJ20}, estimated the mass of the BH as $9.1^{+1.8}_{-1.2}$ $M_{\odot}$
from spectral analysis with the physically motivated TCAF model. The~spin parameter was estimated as $a=0.78^{+0.04}_{-0.04}$, and the inclination angle of
the inner disc was estimated to be $i=29.2^{+0.3}_{-0.5}$ from reflection spectroscopy using NuSTAR observations~\citep{Jia2022}. The~reflection was found
to be from a high-density accretion disk~\citep{Chakraborty2021}. During~the first outburst, type-B QPO was observed from NICER observations
\citep{Belloni2020, Zhang2021}. A time lag between different energy bands was observed during the first outburst \citep{Weng2021}. 

Although the first outburst of MAXI J1348-630 has been studied extensively, the~second outburst has not been studied in detail. In~this paper, we study the
evolution of the spectral and timing properties of the second outburst in detail using data obtained from Swift, MAXI, NICER, NuSTAR and AstroSat
observatories. The~paper is organized as follows. In~Sections 
 \ref{sect:data} and \ref{sect:analysis}, we discuss data reduction and analysis methods, respectively. In~Section \ref{sect:Results}, we present
our results. In~Section \ref{sect:discussion}, we discuss findings and draw our~conclusions.

\section{Data~Reduction}
\label{sect:data}

We studied the BHC~MAXI J1348-630 during its second outburst (2019 May to 2019 August). The~outburst continued for two and a half months. 
In our analysis, we used the data from Swift, NICER, NuSTAR and MAXI for spectral analysis. A~total of 21 observations were used for the 
spectral study. Out of these 21 observations, 8 are combined Swift/XRT and MAXI/GSC, 2 are Swift/XRT-only, 8 are combined NICER and MAXI/GSC, and 3 are NuSTAR-only (check Table~\ref{tab1}). Furthermore, for~timing analysis, we used AstroSat/LAXPC and NICER~data.
\begin{table}[H]
\caption{Log of Swift, NICER and NuSTAR observations of the transient BHC MAXI~J1348-630.\label{tab1}}
\newcolumntype{C}{>{\centering\arraybackslash}X}
\begin{tabularx}{\textwidth}{CCcCcC}
\toprule
\textbf{ID}   &  \textbf{Obs. ID}    &   \textbf{Satellite/Instrument}  &   \textbf{MJD}    &   \textbf{Date of Obs.}  &  \textbf{Exposure}\\
     &             &                         &          &  \textbf{YYYY-MM-DD}     &   \textbf{(ks)} \\
\textbf{(1)}  &    \textbf{(2)}      &          \textbf{(3)}            &   \textbf{(4)}    &   \textbf{(5)}           &   \textbf{(6)}  \\
\midrule
X1   & 00011107040  &   XRT+GSC      & 58,630.70 & 2019-05-27     &  1.02\\ 
X2   & 00011107041  &   XRT+GSC      & 58,633.39 & 2019-05-30     &  0.86\\ 
X3   & 00011107042  &   XRT+GSC      & 58,639.67 & 2019-06-05     &  1.00\\ 
X4   & 00011107043  &   XRT+GSC      & 58,647.82 & 2019-06-13     &  1.07\\
X5   & 00011107044  &   XRT+GSC      & 58,650.84 & 2019-06-16     &  1.00\\
X6   & 00011107045  &   XRT          & 58,685.02 & 2019-07-21     &  0.93\\
X7   & 00011107046  &   XRT+GSC      & 58,689.63 & 2019-07-25     &  2.01\\
X8   & 00011107047  &   XRT+GSC      & 58,690.23 & 2019-07-26     &  1.60\\
X9   & 00011107048  &   XRT          & 58,699.84 & 2019-08-04     &  1.00\\
X10  & 00011107049  &   XRT+GSC      & 58,706.42 & 2019-08-11     &  0.90\\
\midrule
NI1   & 2200530143  &   NICER+GSC    & 58,634.04 & 2019-05-31     &  1.37\\
NI2   & 2200530144  &   NICER+GSC    & 58,637.81 & 2019-06-03     &  1.11\\
NI3   & 2200530170  &   NICER+GSC    & 58,675.91 & 2019-07-11     &  0.49\\
NI4   & 2200530172  &   NICER+GSC    & 58,678.44 & 2019-07-14     &  1.35\\
NI5   & 2200530175  &   NICER+GSC    & 58,681.43 & 2019-07-17     &  1.23\\
NI6   & 2200530185  &   NICER+GSC    & 58,691.62 & 2019-07-27     &  1.70\\
NI7   & 2200530187  &   NICER+GSC    & 58,693.49 & 2019-07-29     &  0.78\\
NI8   & 2200530190  &   NICER+GSC    & 58,696.14 & 2019-08-01     &  0.71\\
\midrule
NU1   & 80502304002 &   NuSTAR       & 58,655.60 & 2019-06-21     &  13.78\\ 
NU2   & 80502304004 &   NuSTAR       & 58,660.73 & 2019-06-26     &  15.37\\ 
NU3   & 80502304006 &   NuSTAR       & 58,672.61 & 2019-07-08     &  17.18\\ 
\bottomrule
\end{tabularx}
\end{table}

\subsection{Swift/XRT}

First, we used the 
 \textsc{xrtpipeline}\endnote{https://www.swift.ac.uk/analysis/xrt, accessed on 15 January 2022}
 command to generate cleaned level-2 event files 
from the level-1 data files. Using the {\tt XSELECT} task, a circular region of radius 30~arcseconds was chosen around the source location to generate 
the source region file. We also chose a background region of radius 30 arcseconds away from the source to produce the background region file. 
Using the region files, we extracted spectra for both the source and the background. With~the help of the tool {\tt XRTMKARF}, we created corresponding 
ARFs. We obtained the appropriate RMFs from the CALDB. Using the {\tt grppha} task, we re-binned the spectra with a minimum of 
20 counts/bin.  

\subsection{NICER}

NICER has unprecedented spectral and timing resolutions of $\sim$85 eV at 1~keV and $\sim$100 nanoseconds, respectively. We used the latest 
calibration files (20210707) for data reduction. To~analyze the NICER data, we first processed the data with the \textsc{nicerl2} 
script, which runs a standard pipeline and produces cleaned level-2 event files with the use of standard calibration. Then, we ran the command 
{\tt barycorr} to produce the barycenter-corrected cleaned event file. The~barycentered-corrected cleaned event files were used to extract 
the light curve and spectrum in the {\tt XSELECT} environment. The~{\tt nibackgen3C50} tool was used to produce the background spectra 
corresponding to each of the observation IDs. The~spectra were re-binned with a minimum of 20 counts per bin with the use of {\tt grppha}.
To search for QPOs, we extracted the 0.01 s binned light curves from the barycenter-corrected cleaned event file using {\tt xselect}.


\subsection{NuSTAR}

We used NuSTAR/FPMA data in the 4--78~keV energy band  to get broad energy information. Data reduction was done using the NuSTAR 
data analysis software \textsc{NuSTARDAS}. First, the \textsc{nupipeline} command was run to get the stage-II data for the Focal Plane 
Module FPMA. A~circular region of 80 arcseconds was chosen at the source location to generate the source region file. We also chose a region 
of 80 arcseconds away from the source to generate the background region file. We generated spectra, RMF and ARF files using {\tt nuproduct}. The~extracted spectra were re-binned to have at least 30 counts per bin with the tool {\tt grppha}.

\subsection{MAXI/GSC}

We used the MAXI on-demand process web tool to generate 6--20~keV MAXI/GSC spectra using the process mentioned in Matsuoka~et~al., 2009~\citep{Matsuoka09}. 
The MAXI/GSC spectrum files are available from (\url{http://maxi.riken.jp/mxondem}, accessed on 15 January 2022
). We downloaded the MAXI/GSC spectral observations that were simultaneous/quasi-simultaneous with those of XRT and NICER. One-day-averaged light curves in different energy ranges available from
(\url{http://maxi.riken.jp/top/lc.html}, accessed on 11 January 2022) were used to observe the variation of~flux.

\subsection{AstroSat/LAXPC}

We used publicly available code from the LAXPC Software website to produce cleaned level-2 data from the level-1 data files from  
LAXPC20. For~each observation, we initially ran the `{\tt laxpcl1.f}' program to process  multiple  orbits  of  level-1  data. 
This program outputs event files, light curves, spectra and GTI files in both ASCII and FITS formats. 
Then, the `{\tt backshiftv3.f}' program was used to apply background correction to the light curve files. For~each observation, 
we initially ran the program `{\tt laxpcl1.f}' with a time bin of 1 s and the full range of anodes and channels. The~program produced 
output files along with the {\tt lxp1level2.gti}, which was moved to the file {\tt gti.inp}. Once the {\tt gti.inp} file was prepared, 
we extracted the 0.01 s binned light curve by running the program `{\tt laxpcl1.f}’ again with the full range of anodes and channels. 
We used the 0.01 s time binned light curves (3--80~keV) to search for the~QPOs. 

\section{Data~Analysis}
\label{sect:analysis}

\subsection{Temporal~Analysis}
Archival data from MAXI/GSC\endnote{http://maxi.riken.jp/top/lc.html, accessed on 11 January 2022}
(2--10~keV) and SWIFT/BAT\endnote{https://swift.gsfc.nasa.gov/results/transients/, accessed on 11 January 2022} 
(15--50~keV) were used to observe the variation of fluxes during the outburst. 
The one-day-averaged light curves were converted into $Crab$ units using proper conversion factors. The~Crab conversion factor for GSC 
(2--10~keV) data is 2.82~photons~cm$^{-2}$~s$^{-1}$, and 
 for BAT, the Crab conversion factor is 0.218~counts~cm$^{-2}$~s$^{-1}$. To~search 
for the QPOs, we generated the white-noise-subtracted power density spectra (PDS) using 0.01 s time binned light curves of AstroSat 
(3--80~keV) and NICER (1--10~keV) data. The~{\tt powspec} task of the \textsc{XRONOS} software package was used to generate the PDS from 
the 0.01 s time binned light curves. Each light curve was divided into 8192 intervals, and a PDS for each interval was generated; these
were normalized so that their integral gave the squared rms fractional variability. To~obtain the final PDS, all individual PDSs were 
then averaged. We used geometrical re-binning constants of ${-1.02}$ or $-1.05$ as needed on the final PDSs. We used a multiple Lorentzian model to fit the PDS. From~this, we obtained the fitted values of QPO frequencies ($\nu_{QPO}$), width ($\triangle \nu$), Q-value 
($Q=\nu_{QPO}/\triangle\nu$) and RMS ($\%$) amplitude. With~the help of these values, we classified the nature of the~QPOs.

\subsection{Spectral~Analysis}
For spectral analysis, we used XRT spectra from 1--8~keV, GSC from 6--20~keV, NICER from 1--10~keV and NuSTAR from 4--78~keV energy ranges.
For our spectral study, we used 21 total observations. We analyzed simultaneous/quasi-simultaneous XRT and GSC data from 8~observations 
in the energy range of 1--20~keV. Since no simultaneous GSC data were found, 2 XRT-only observations were used in the energy range
of 1--8~keV. Combined GSC and NICER data were studied in the 1--20~keV energy range from 8 observations. We also studied three NuSTAR 
observations in the 4--78~keV energy range (see Table~\ref{tab1}). First, the 1--20~keV and 1--8~keV spectra were fitted with the absorbed 
powerlaw model. {We also fitted 4--78~keV NuSTAR spectra with the absorbed powerlaw model; however, a~sign of reflection was seen in 
the residuals near 10~keV. We used convolution model \textsc{reflect} for the reprocessed emission with the powerlaw model. The~model 
\textsc{reflect} is a convolution model for reflection from neutral material in accordance with the approach of}  Magdziarz and Zdziarski (1995)~\citep{Magdziarz1995}.
{The model has five parameters: reflection scaling factor ($rel_{refl}$), redshift (z),
abundance of elements heavier than He relative to solar abundances, iron abundance and~cosine of inclination angle (cosIncl). While fitting, we froze the
iron abundance and heavy-element abundance to the solar value (i.e., 1) and redshift as 0.0 
in the \textsc{reflect} model. We allowed the relative reflection ($rel_{refl}$) and~inclination angle of the system (as cosIncl) to vary.} The model read
in {\tt XSPEC} as \textsc{reflect}*powerlaw.  We also added a Gaussian at 6.4~keV to incorporate the Fe K$\alpha$ line.

Next, we used the physical model TCAF for spectral analysis. The~1--20~keV and 1--8~keV spectra were fitted with the absorbed TCAF model.
The 4--78~keV NuSTAR spectra were fitted with the \textsc{reflect}*TCAF model. Similar to the powerlaw model, we used \textsc{reflect} for the
reprocessed emission in the TCAF model. A~Gaussian line at 6.4~keV was also added to incorporate the Fe K$\alpha$ line. For~all of the 21 observations, we used the TBabs model to account for absorption in the interstellar
medium. We used a 1\% systematic error for the Swift, combined Swift/XRT and MAXI/GSC and combined NICER and MAXI/GSC~spectra.

\section{Results}
\label{sect:Results}
During this second outburst of the BHC MAXI~J1348-630 in 2019, the~source was in the active phase for around two and a half months. 
We have studied the temporal and spectral properties of the BHC during this outburst. Here, we present the results of our~analysis.

\subsection{Temporal~Properties}
\unskip

\subsubsection{Outburst~Profiles}
\begin{figure}[H]
\includegraphics[width=11truecm,angle=0]{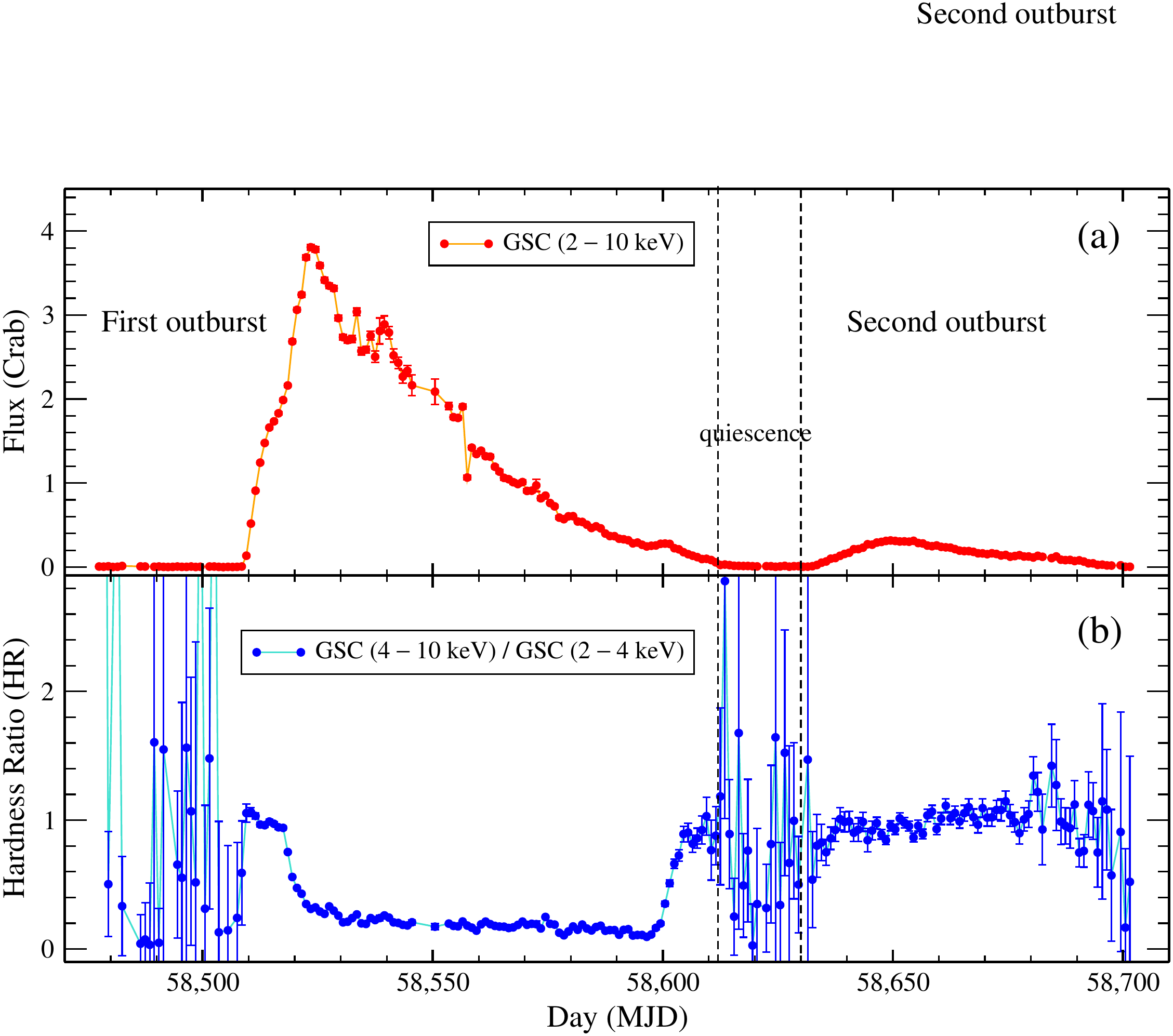}
\caption{Variation of (\textbf{a}) 2--10~keV MAXI/GSC flux and (\textbf{b}) hardness ratio (HR) of 4--10~keV to 2--4~keV MAXI/GSC fluxes of the BHC MAXI J1348-630 during the first and second~outbursts.\label{fig1}}


\end{figure}

Figure~\ref{fig1} {shows the outburst profile, 2--10~keV MAXI/GSC flux (a) and hardness ratio (HR) diagram (b) of BHC MAXI~J1348-630 during the first and second outbursts. From~the figure, we can see that the source went to quiescence (for $\sim$20~days)
after the completion of its first outburst (MJD$\sim$58,500 to MJD$\sim$58,610) and again rebrightened from MJD$\sim$58,630. The~second
outburst was a mini outburst compared to the first one.} 

Figure~\ref{fig2} shows the light curves in different energy ranges and the hardness ratios of the second outburst in 
2019 of BHC MAXI~J1348-630. Panel (a) of Figure~\ref{fig2} shows the~variation of 15--50~keV Swift/BAT flux (online blue) and 
2--10~keV MAXI/GSC flux (online red). Panel (c) shows the variation of MAXI/GSC flux in two different energy bands: 4--10~keV (online blue) and 2--4~keV (online red). The~outburst started on MJD$\sim$58,630 (27 May 2019) 
and returned to quiescence after MJD$\sim$58,707 (12 August 2019). According to the variation of flux or count rates in light curve 
profiles, outbursts are generally divided into two categories: fast-rise slow-decay (FRSD) and slow-rise slow-decay (SRSD) 
\citep{DD10}. From~light curve profiles, this outburst can be characterized as slow-rise slow-decay (SRSD). 
As seen in panel (a), both the Swift/BAT flux (15--50~keV) and the total MAXI/GSC flux (2--10~keV) started to increase 
slowly from MJD$\sim$58,630. At~MJD$\sim$58,650, both the fluxes reached their maximum values, and afterward started to decrease 
slowly. The~fluxes reached quiescence after MJD$\sim$58701. Hard (BAT) flux was dominant over soft (GSC) flux during the 
entire outburst. Panel (c) shows that flux is low in both energy bands at the start of the outburst.
Here, the fluxes gained maximum values near MJD$\sim$58,650. Afterwards, both fluxes declined slowly till the end of 
the~outburst.

\subsubsection{Hardness~Ratio}
Hardness ratio (HR) is defined as the ratio of hard X-ray flux to soft X-ray flux. Variation of the HR provides us with 
a rough idea of the evolution of the flow dynamics of the source, as it is believed that the origin of soft X-ray flux is thermal and 
hard X-ray flux is non-thermal. The~HR is generally high in harder states (HS and HIMS), since high-energy photons
dominate over soft photons in harder states. The~HR is low in softer states (SIMS and SS) as the reverse condition takes 
place in softer states. For~a complete outburst, the~HR remains high at the beginning of the outburst since the BH remains 
in the hard state (rising). As~time progresses the HR gradually decreases in the rising intermediate states (HIMS and SIMS). 
In the SS, HR has a low value and remains almost constant. The~HR gradually increases in the declining intermediate states 
and becomes roughly constant at a high value in the HS (declining).

\begin{figure}[H]
\includegraphics[width=10.5 cm]{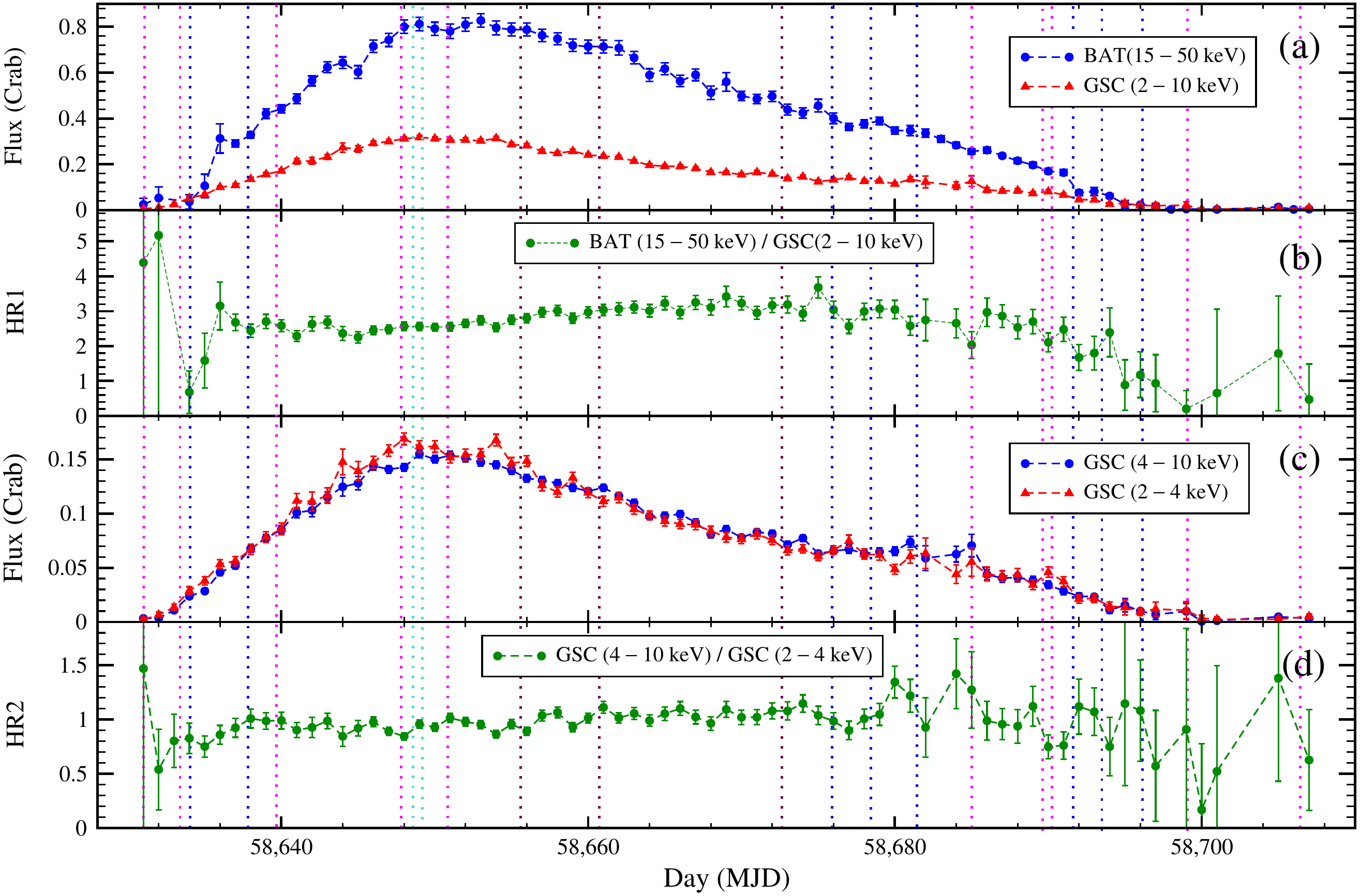}
\caption{The 
 variation of (\textbf{a}) Swift/BAT flux in 15--50~keV (blue) and 2--10~keV MAXI/GSC (red) fluxes; (\textbf{b}) hardness ratio (HR1)
of BAT (15--50~keV) and GSC (2--10~keV) fluxes; (\textbf{c}) MAXI/GSC fluxes in 4--10~keV (blue) and 2--4~keV (red); (\textbf{d}) hardness ratio (HR2)
of 4--10~keV to 2--4~keV MAXI/GSC fluxes. Vertical lines indicates dates of the observation IDs used in the work.
Magenta indicates Swift/XRT, blue indicates NICER, and maroon indicates NuSTAR observation IDs. Turquoise 
indicates the epoch of the AstroSat observations for timing analysis. 
} \label{fig2}
\end{figure}

In Figure~\ref{fig2}, we show two HRs in (HR1 and HR2 in (b) and (d), respectively): (b) shows the ratio of 15--50~keV Swift/BAT flux to 
2--10~keV total MAXI/GSC flux, and (d) shows the ratio of 4--10~keV MAXI/GSC hard flux to 2--4~keV MAXI/GSC soft flux. 
In HR1 (b) we notice that at the beginning of the outburst, the HR has a value of$\sim$2.7 near MJD$\sim$58,639. The~value 
slowly increases to 3.42 around MJD$\sim$58,669, and after that it slowly decreases till the end of the outburst.
In HR2 (d), at the beginning of the outburst, the~HR was around$\sim$1, and it stayed there until 
MJD$\sim$58,679. After~that, the HR slightly increased to a value of 1.42 at MJD 58,684 and decreased afterwards till the end 
of the outburst. There, we did not find any signatures of state transition in either of the HRs (HR1 and HR2) as could be seen for 
a complete or normal outburst of classical BH sources. The~HRs varied a little around a certain value throughout the outburst. 
From the hardness ratio, we can roughly say that the source did not go to softer states; rather, it remained in harder states throughout 
the outburst. To~confirm this, we need to perform spectral~analysis.



\subsubsection{Power Density~Spectra}
\begin{figure}[H]
\includegraphics[width=11truecm,angle=0]{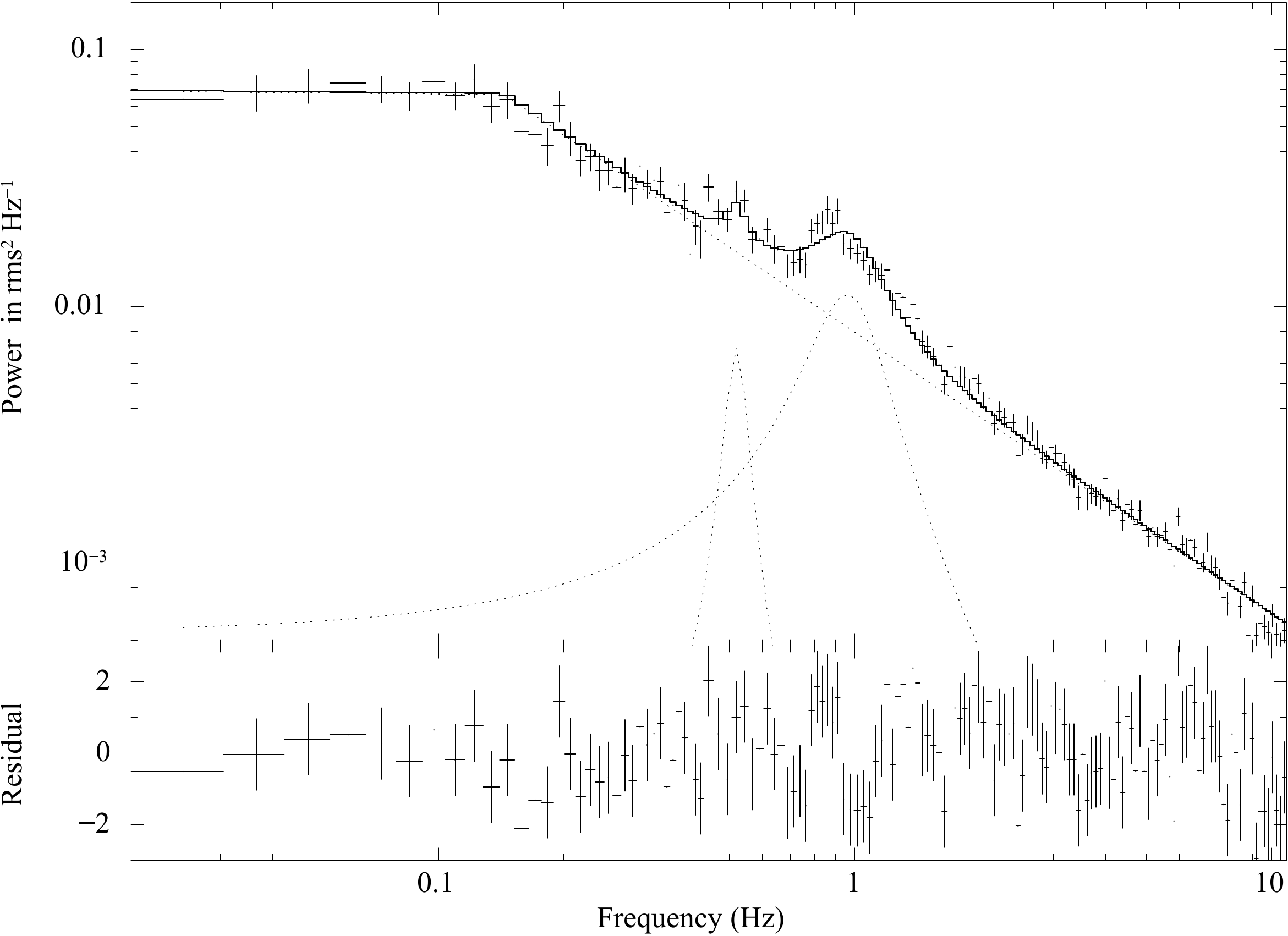}
\caption{Continuum-fitted power density spectrum using 0.01 s time binned 3--80~keV AstroSat/LAXPC light curve from orbit 20064 of 
observation ID= T03\_120T01\_9000002990 with a QPO of frequency $0.96\pm0.01$~Hz along with a sub-harmonic at $0.52\pm0.01$~Hz .\label{fig3}}
\end{figure}

We studied the power-density spectra (PDS) generated with the 0.01 s light curves of AstroSat/LAXPC (3--80~keV) and NICER (1--10~keV). 
We observed QPOs only on two days, 14 June 2019 and 15 June 2019, 
in AstroSat/LAXPC data. A~QPO of centroid frequency $0.96\pm0.01$~Hz was 
observed on 14 June 2019
(MJD~=~58,648.6) with a Q-value of $2.23\pm0.21$ and $6.87\pm0.4\%$ rms. Figure~\ref{fig3} is a~continuum-fitted 
We fitted the PDS with a broken powerlaw and two Lorentzian models. {In the PDS, it can be seen
that along with the primary QPO at $0.96\pm0.01$~Hz, there is one more QPO feature at $0.52\pm0.01$~Hz. The~Q-value of this weaker QPO is
$8.67\pm1.45$, and the rms value is $0.80\pm0.1\%$. This is a sub-harmonic of the primary QPO observed at $0.96\pm0.01$~Hz.} Another QPO was
observed on 15 June 2019 (MJD~=~58,649.2) with a centroid frequency of $0.95\pm0.02$~Hz and a Q-value of $2.57\pm0.42$ and $12.0\pm1.1\%$ rms.
{ We also calculated the value of the characteristic frequency with the formula $\nu_{max}=\sqrt{\nu_{0}^{2}+(\Delta/2)^{2}}$,
where $\nu_{0}$ is the centroid frequency, and $\Delta$ is the FWHM of the Lorentzian}~\citep{Belloni1997, Belloni2002, VanStraaten2002}. 
{For 14 June 2019, the~value of the characteristic frequency of the primary QPO is $0.98\pm0.02$~Hz, and on the next day it is observed 
at $0.98\pm0.03$~Hz}. Note, we did not observe any prominent QPO nature in the NICER data; this {might be due to lower effective area and lower 
energy band compared to AstroSat/LAXPC.}
\unskip

\subsection{Spectral~Properties}
To get an idea about the source and its evolution during the outburst, the~study of the spectral properties is very important. 
As discussed earlier, we used 1--20~keV combined data from Swift/XRT and MAXI/GSC (8 OBSIDs), 1--20~keV NICER and MAXI/GSC (8 OBSIDs), 
1--8~keV Swift/XRT (2 OBSIDs) and 4--78~keV NuSTAR data (3 OBSIDs) for the spectral analysis of BHC MAXI~J1348-630 during this outburst. First, we fitted all 15 observations of Swift/XRT, (Swift/XRT+MAXI/GSC) data and (NICER+MAXI/GSC) data with the powerlaw (PL) 
model. The~fitted parameters are shown in Table~\ref{tab2}. Since there were hints of presence of reflection in the NuSTAR spectra (after fitting 
spectra with only the PL or TCAF model), we fitted the three publicly available NuSTAR data with the \textsc{reflect}*powerlaw
model along with a Gaussian line. The~values of the fitted parameters are shown in Table~\ref{tab3}.

The PL model gives us a rough idea about the spectral states. To~know about the variation of the accretion flow parameters during 
the outburst, we fitted all the observations with the TCAF model also. {Initially, we fitted the spectra with the TCAF model keeping the BH mass ($M_{BH}$) free. From~each 
spectral fit, we obtained the best-fitted value of $M_{BH}$. The~$M_{BH}$ values varied between 7.9 $M_\odot$ and 10.3 $M_\odot$.
We averaged these best-fitted mass values to get 9.1 $M_\odot$. Then, we kept $M_{BH}$
frozen at this value and refitted all the spectra to obtain the final result. The~preliminary values of the fitted parameters of the 1--20~keV XRT+GSC data, 
1--8~keV XRT data and the 1--20~keV NICER+GSC spectra are mentioned in Table~\ref{ta1}.  The~final results are shown in }Table~\ref{tab2}. The~NuSTAR data is fitted 
with the TCAF model along with \textsc{reflect} model (to account for the reflection) and a Gaussian line to incorporate the Fe K$\alpha$ emission line. 
{The preliminary values of the fitted parameters are shown in Table~\ref{ta2} (where we kept the $M_{BH}$ as a free parameter). The~final fitted parameters 
are mentioned in} Table~\ref{tab4} {(where we froze the $M_{BH}$ at 9.1 $M_\odot$)}.

\begin{table}[H]
\caption{Model-fitted parameters for Swift/XRT, combined NICER+MAXI/GSC and Swift/XRT+MAXI/GSC spectra (mass, $M_{BH}$, is frozen at 9.1 $M_{\odot}$ for TCAF model).\label{tab2}}
\begin{adjustwidth}{-\extralength}{0cm}
		\newcolumntype{C}{>{\centering\arraybackslash}X}
		\begin{tabularx}{\fulllength}{CCCCCCCCCC}
\toprule
\textbf{ID\boldmath{$^{[1]}$}} & \boldmath{$n_H^{[2]}$} & \boldmath{$\Gamma^{[3]}$} & \boldmath{$flux^{[3]}$} & \boldmath{${\chi}^2$\textbf{/dof}$^{[5]}$} & \boldmath{${\dot m}_d$$^{[4]}$} & \boldmath{${\dot m}_h$$^{[4]}$} & \textbf{R\boldmath{$^{[4]}$}} & \boldmath{$X_s$$^{[4]}$} & \boldmath{${\chi}^2$\textbf{/dof}$^{[5]}$} \\
 \textbf{(1)}        & \textbf{(2)}      & \textbf{(3)}     & \textbf{(4)}    &      \textbf{(5)}           & \textbf{(6)}     & \textbf{(7)}              & \textbf{(8)}            & \textbf{(9)}      &  \textbf{(10)}     \\
\midrule
X1  & $0.64^{\pm0.20}$ & $1.59^{\pm0.07}$ & $0.36^{\pm0.04}$ & 22/27   & $1.20^{\pm0.08}$ & $0.24^{\pm0.02}$ & $3.41^{\pm0.07}$ & $239^{\pm12}$  & 19/24\\
X2  & $0.78^{\pm0.14}$ & $1.62^{\pm0.11}$ & $3.67^{\pm0.25}$ & 199/197 & $1.26^{\pm0.10}$ & $0.32^{\pm0.02}$ & $3.50^{\pm0.02}$ & $217^{\pm5}$   & 197/194\\
NI1 & $0.66^{\pm0.05}$ & $1.64^{\pm0.01}$ & $6.64^{\pm0.03}$ & 764/709 & $1.50^{\pm0.10}$ & $0.36^{\pm0.02}$ & $3.51^{\pm0.002}$ & $210^{\pm2}$  & 777/706\\
NI2 & $0.51^{\pm0.03}$ & $1.66^{\pm0.01}$ & $25.2^{\pm0.10}$ & 924/839 & $1.52^{\pm0.04}$ & $0.43^{\pm0.03}$ & $3.57^{\pm0.008}$ & $205^{\pm8}$  & 996/836\\
X3  & $0.65^{\pm0.09}$ & $1.68^{\pm0.06}$ & $33.7^{\pm1.50}$ & 721/622 & $1.63^{\pm0.01}$ & $0.45^{\pm0.02}$ & $3.41^{\pm0.01}$ & $195^{\pm14}$  & 725/619\\
X4  & $0.62^{\pm0.05}$ & $1.74^{\pm0.03}$ & $64.5^{\pm0.50}$ & 959/686 & $1.75^{\pm0.01}$ & $0.77^{\pm0.01}$ & $2.82^{\pm0.01}$ & $185^{\pm4}$ & 932/683\\
X5  & $0.50^{\pm0.13}$ & $1.78^{\pm0.02}$ & $63.0^{\pm1.00}$ & 707/627 & $2.32^{\pm0.04}$ & $0.70^{\pm0.01}$ & $2.71^{\pm0.01}$ & $125^{\pm4}$  & 725/624\\
NI3 & $0.51^{\pm0.03}$ & $1.65^{\pm0.01}$ & $31.0^{\pm0.10}$ & 788/779 & $1.48^{\pm0.07}$ & $0.59^{\pm0.02}$ & $2.77^{\pm0.01}$ & $127^{\pm2}$  & 816/776\\
NI4 & $0.53^{\pm0.02}$ & $1.65^{\pm0.01}$ & $28.2^{\pm0.20}$ & 1037/872& $1.51^{\pm0.07}$ & $0.50^{\pm0.03}$ & $2.77^{\pm0.03}$ & $124^{\pm8}$  & 1099/869\\
NI5 & $0.55^{\pm0.03}$ & $1.66^{\pm0.01}$ & $25.9^{\pm0.20}$ & 979/841 & $1.45^{\pm0.03}$ & $0.40^{\pm0.01}$ & $2.73^{\pm0.01}$ & $127^{\pm2}$  & 964/838\\
X6  & $0.79^{\pm0.20}$ & $1.66^{\pm0.06}$ & $19.6^{\pm0.40}$ & 561/522 & $1.37^{\pm0.08}$ & $0.31^{\pm0.04}$ & $2.76^{\pm0.05}$ & $132^{\pm4}$  & 570/519\\
X7  & $0.50^{\pm0.29}$ & $1.66^{\pm0.17}$ & $19.0^{\pm3.30}$ & 57/48   & $1.12^{\pm0.20}$ & $0.39^{\pm0.07}$ & $2.82^{\pm0.03}$ & $152^{\pm20}$ & 55/45\\
X8  & $0.93^{\pm0.25}$ & $1.68^{\pm0.03}$ & $14.1^{\pm0.20}$ & 596/489 & $1.15^{\pm0.04}$ & $0.31^{\pm0.05}$ & $2.80^{\pm0.01}$ & $154^{\pm2}$ & 596/486\\
NI6 & $0.61^{\pm0.01}$ & $1.64^{\pm0.02}$ & $11.7^{\pm0.20}$ & 771/801 & $1.13^{\pm0.01}$ & $0.35^{\pm0.03}$ & $3.50^{\pm0.05}$ & $155^{\pm5}$ & 899/798\\
NI7 & $0.60^{\pm0.02}$ & $1.63^{\pm0.01}$ & $6.91^{\pm0.03}$ & 612/637 & $1.09^{\pm0.03}$ & $0.31^{\pm0.05}$ & $3.54^{\pm0.02}$ & $160^{\pm4}$ & 639/634\\
NI8 & $0.63^{\pm0.04}$ & $1.62^{\pm0.01}$ & $3.11^{\pm0.05}$ & 469/517 & $1.00^{\pm0.02}$ & $0.30^{\pm0.01}$ & $3.53^{\pm0.01}$ & $176^{\pm8}$ & 468/514\\
X9  & $0.83^{\pm0.13}$ & $1.63^{\pm0.14}$ & $0.71^{\pm0.04}$ & 55/50   & $1.00^{\pm0.08}$ & $0.23^{\pm0.01}$ & $3.61^{\pm0.05}$ & $185^{\pm4}$  & 56/47\\
X10 & $0.84^{\pm0.04}$ & $1.62^{\pm0.14}$ & $0.72^{\pm0.04}$ & 64/52   & $1.00^{\pm0.08}$ & $0.22^{\pm0.19}$ & $3.61^{\pm0.05}$ & $188^{\pm4}$  & 65/49\\
\bottomrule
\end{tabularx}
\end{adjustwidth}
\noindent{\footnotesize{} {$^{[1]}$ ID of the observed dates as mentioned in Table~\ref{tab1} (Col. 1).}
{$^{[2]}$ Model-fitted value of hydrogen column density ($n_H$) in $10^{22}$ atoms per cm$^{-2}$ (Col. 2).}
{$^{[3]}$ PL model-fitted photon index ($\Gamma$) in Col. 3.}
{$^{[3]}$ PL model-fitted flux in Col. 4 in $10^{-10}$ order}
{$^{[4]}$ TCAF model-fitted parameters: disk rate (${\dot m}_d$ in Eddington rate ${\dot M}_{Edd}$) in $10^{-3}$ order,} 
{halo rate (${\dot m}_h$ in ${\dot M}_{Edd}$), compression ratio ($R$) and shock location ($X_s$ in Schwarzschild radius $r_s$)};
{Cols. 6--9, respectively.} 
{$^{[5]}$ PL and TCAF model-fitted ${\chi}^2_{red}$ values; Cols. 5 and 10, respectively, as}
{${\chi}^2/dof$, where `dof' represents degrees of freedom.}
{Note: We present average values of 90\% confidence $\pm$ parameter error values, which are obtained using `err' task in XSPEC.}}
\end{table}
\vspace{-10pt} 

\begin{table}[H]
\tablesize{\fontsize{8}{8}\selectfont}
\caption{Fitted parameters for NuSTAR data with PL/\textsc{reflect}*PL model along with a Gaussian~line.}\label{tab3}
\begin{adjustwidth}{-\extralength}{0cm}
		\newcolumntype{C}{>{\centering\arraybackslash}X}
		\begin{tabularx}{\fulllength}{cCCCCCCCCCCC}
\toprule
               &\textbf{ID\boldmath{$^{[1]}$}}& \boldmath{$n_H$$^{[2]}$} & \boldmath{$rel_{refl}$$^{[3]}$} & \boldmath{$cosIncl$$^{[3]}$} & \boldmath{$\Gamma^{[4]}$} & \boldmath{$flux^{[4]}$} & \boldmath{$norm^{[4]}$} & \textbf{lineE\boldmath{$^{[5]}$}} & \boldmath{$\sigma^{[5]}$} & \textbf{norm\boldmath{$^{[5]}$}} & \textbf{\boldmath{${\chi}^2$}/dof$^{[6]}$} \\
               & \textbf{(1)}      &    \textbf{(2)}        &        \textbf{(3)}           &    \textbf{(4)}            &     \textbf{(5)}        &      \textbf{(6)}     &       \textbf{(7)}      &     \textbf{(8)}       &    \textbf{(9)} &      \textbf{(10)}    & \textbf{(11)}  \\
\midrule
Model 1:       &NU1& $1.01^{\pm0.12}$ &             &              & $1.63^{\pm0.01}$  & $59.5^{\pm0.30}$ & $1.46^{\pm0.01}$ & $6.20^{\pm0.19}$ & $0.49^{\pm0.11}$ & $2^{\pm0.5}$ & 2494/1500\\
TBabs*(PL      &NU2& $1.04^{\pm0.12}$ &             &              & $1.60^{\pm0.01}$  & $52.4^{\pm0.10}$ & $1.17^{\pm0.01}$ & $6.20^{\pm0.17}$ & $0.51^{\pm0.10}$ & $2^{\pm0.5}$ & 2689/1501\\
+Gaussian)     &NU3& $1.01^{\pm0.13}$ &             &              & $1.58^{\pm0.01}$  & $34.3^{\pm0.20}$ & $0.74^{\pm0.01}$ & $6.46^{\pm0.05}$ & $0.22^{\pm0.01}$ & $1^{\pm0.1}$ & 2248/1439\\
\midrule
Model 2:       &NU1& $0.74^{\pm0.14}$ &$0.23^{\pm0.05}$&$0.86^{\pm0.15}$ & $1.70^{\pm0.01}$ & $62.2^{\pm0.20}$ & $1.55^{\pm0.03}$ &  $6.40^{\pm0.07}$ & $0.75^{\pm0.08}$ & $4^{\pm0.6}$ & 1594/1498\\
TBabs*(reflect &NU2& $0.50^{\pm0.26}$ &$0.26^{\pm0.05}$&$0.70^{\pm0.14}$ & $1.67^{\pm0.01}$ & $53.6^{\pm0.30}$ & $1.28^{\pm0.02}$ &  $6.31^{\pm0.08}$ & $0.65^{\pm0.08}$ & $3^{\pm0.5}$ & 1553/1499\\
*PL+Gaussian)  &NU3& $0.74^{\pm0.14}$ &$0.26^{\pm0.06}$&$0.70^{\pm0.17}$ & $1.66^{\pm0.01}$ & $34.8^{\pm0.20}$ & $0.84^{\pm0.01}$ &  $6.51^{\pm0.05}$ & $0.25^{\pm0.06}$ & $1^{\pm0.2}$ & 1464/1437\\
\bottomrule
\end{tabularx}
\end{adjustwidth}
\noindent{\footnotesize
{$^{[1]}$ ID of the observed dates as described in Table~\ref{tab1} (Col. 1).}
{$^{[2]}$ Model-fitted value of hydrogen column density $n_H$ in $10^{22}$ atoms per cm$^{-2}$ (Col. 2).}
{$^{[3]}$ In case of reflect model, reflection scaling factor ($rel_{refl}$) and cosine of inclination}
{angle ($cosIncl$) are mentioned in Cols. (3--4), respectively.} 
{$^{[4]}$ In case of PL model, photon index ($\Gamma$), flux and norm are mentioned in Cols. 5, 6 and 7, respectively.}
{$^{[5]}$ Line energy of the Gaussian line energy (lineE) in~keV, line width (sigma) in~keV}  and
{total photons/cm$^2$/s in the line (norm) in $10^{-3}$ order; Cols. 8--10, respectively.} 
{$^{[6]}$ Model-fitted ${\chi}^2_{red}$; Col. 11 as ${\chi}^2/dof$, where `dof' represents degrees of freedom.}}
\end{table}

\begin{table}[H]
\tablesize{\fontsize{7}{7}\selectfont}
\caption{Fitted parameters for NuSTAR data with TCAF/\textsc{reflect}*TCAF model along with a Gaussian line (mass, $M_{BH}$, is frozen at 9.1 $M_{\odot}$ for TCAF model).}\label{tab4}
\begin{adjustwidth}{-\extralength}{0cm}
		\newcolumntype{C}{>{\centering\arraybackslash}X}
		\begin{tabularx}{\fulllength}{ccCCCCCCCCCCC}
\toprule
&\boldmath{\textbf{ID}$^{[1]}$}  & \boldmath{$n_H$$^{[2]}$} & \boldmath{$rel_{refl}$$^{[3]}$} & \boldmath{$cosIncl$$^{[3]}$} & \boldmath{${\dot m}_d$$^{[4]}$} & \boldmath{${\dot m}_h$$^{[4]}$} & \boldmath{$R$$^{[4]}$} & \boldmath{$X_s$$^{[4]}$} & \boldmath{\textbf{lineE}$^{[5]}$} & \boldmath{\textbf{sigma}$^{[5]}$} & \boldmath{\textbf{norm}$^{[5]}$} & \boldmath{${\chi}^2$\textbf{/dof}$^{[6]}$}  \\
& \textbf{(1)} & \textbf{(2)} &          \textbf{(3) }    &        \textbf{(4)}        &     \textbf{(5)}          &       \textbf{ (6)}      &      \textbf{(7)}         &        \textbf{(8)}       &       \textbf{(9)}        &   \textbf{(10)}      &    \textbf{(11)}  & \textbf{(12)}  \\
\midrule
Model 1:      & NU1 & $0.50^{\pm0.12}$ &&& $1.48^{\pm0.02}$ & $0.61^{\pm0.02}$ & $3.44^{\pm0.10}$ & $134^{\pm1}$ & $6.20^{\pm0.20}$ & $0.80^{\pm0.16}$ & $4^{\pm0.8}$&2113/1497\\
TBabs*(TCAF+  & NU2 & $0.50^{\pm0.12}$ &&& $1.50^{\pm0.10}$ & $0.60^{\pm0.04}$ & $3.65^{\pm0.11}$ & $134^{\pm1}$ & $6.20^{\pm0.21}$ & $0.76^{\pm0.08}$ & $6^{\pm0.7}$&1995/1498\\ 
Gaussian)     & NU3 & $0.50^{\pm0.13}$ &&& $1.04^{\pm0.03}$ & $0.60^{\pm0.03}$ & $3.43^{\pm0.12}$ & $134^{\pm1}$ & $6.20^{\pm0.24}$ & $0.73^{\pm0.10}$ & $3^{\pm0.9}$&1791/1436\\
\midrule
Model 2:         & NU1 & $0.50^{\pm0.23}$ & $0.29^{\pm0.05}$ & $0.86^{\pm0.20}$ & $2.10^{\pm.07}$  & $0.67^{\pm0.01}$ & $2.70^{\pm0.08}$ & $133^{\pm2}$ & $6.33^{\pm0.17}$ & $0.80^{\pm0.25}$ & $5^{\pm0.2}$& 1621/1495\\
TBabs*(reflect*  & NU2 & $0.50^{\pm0.43}$ & $0.32^{\pm0.13}$ & $0.74^{\pm0.21}$ & $2.00^{\pm0.16}$ & $0.62^{\pm0.02}$ & $2.62^{\pm0.01}$ & $130^{\pm5}$ & $6.33^{\pm0.18}$ & $0.69^{\pm0.19}$ & $3^{\pm0.9}$& 1670/1496\\
TCAF+Gaussian)   & NU3 & $0.80^{\pm0.21}$ & $0.33^{\pm0.06}$ & $0.72^{\pm0.17}$ & $1.50^{\pm0.10}$ & $0.60^{\pm0.02}$ & $2.69^{\pm0.01}$ & $125^{\pm9}$ & $6.51^{\pm0.05}$ & $0.22^{\pm0.07}$ & $1^{\pm0.2}$& 1684/1434\\
\bottomrule
\end{tabularx}
	\end{adjustwidth}
\noindent{\footnotesize
{$^{[1]}$ IDs of the observed dates as described in Table~\ref{tab1} (Col. 1).}
{$^{[2]}$ Model-fitted value of hydrogen column density $n_H$ in $10^{22}$ atoms per cm$^{-2}$ (Col. 2).}
{$^{[3]}$ In case of \textsc{reflect}*TCAF+Gaussian model, the~value of reflection scaling factor ($rel_{refl}$)}
{and the cosine of inclination angle ($cosIncl$); Col. 3 and Col. 4, respectively.}
{$^{[4]}$ TCAF model-fitted parameters: disk rate (${\dot m}_d$ in Eddington rate ${\dot M}_{Edd}$) in $10^{-3}$ order,} 
{halo rate (${\dot m}_h$ in ${\dot M}_{Edd}$), compression ratio ($R$) and shock location ($X_s$ in Schwarzschild radius $r_s$);} 
{Cols. 5--8.}
{$^{[5]}$ Line energy of the Gaussian line energy (lineE) in~keV, line width (sigma) in~keV} and
{total photons/cm$^2$/s in the line (norm) in $10^{-3}$ order; Cols. 9, 10 and~11, respectively.}
{$^{[6]}$ Model-fitted ${\chi}^2_{red}$; Col. 12 as ${\chi}^2/dof$, where `dof' represents degrees of freedom.}}
\end{table}

\begin{figure}[H]
\vbox{
\includegraphics[width=6.5truecm,angle=0]{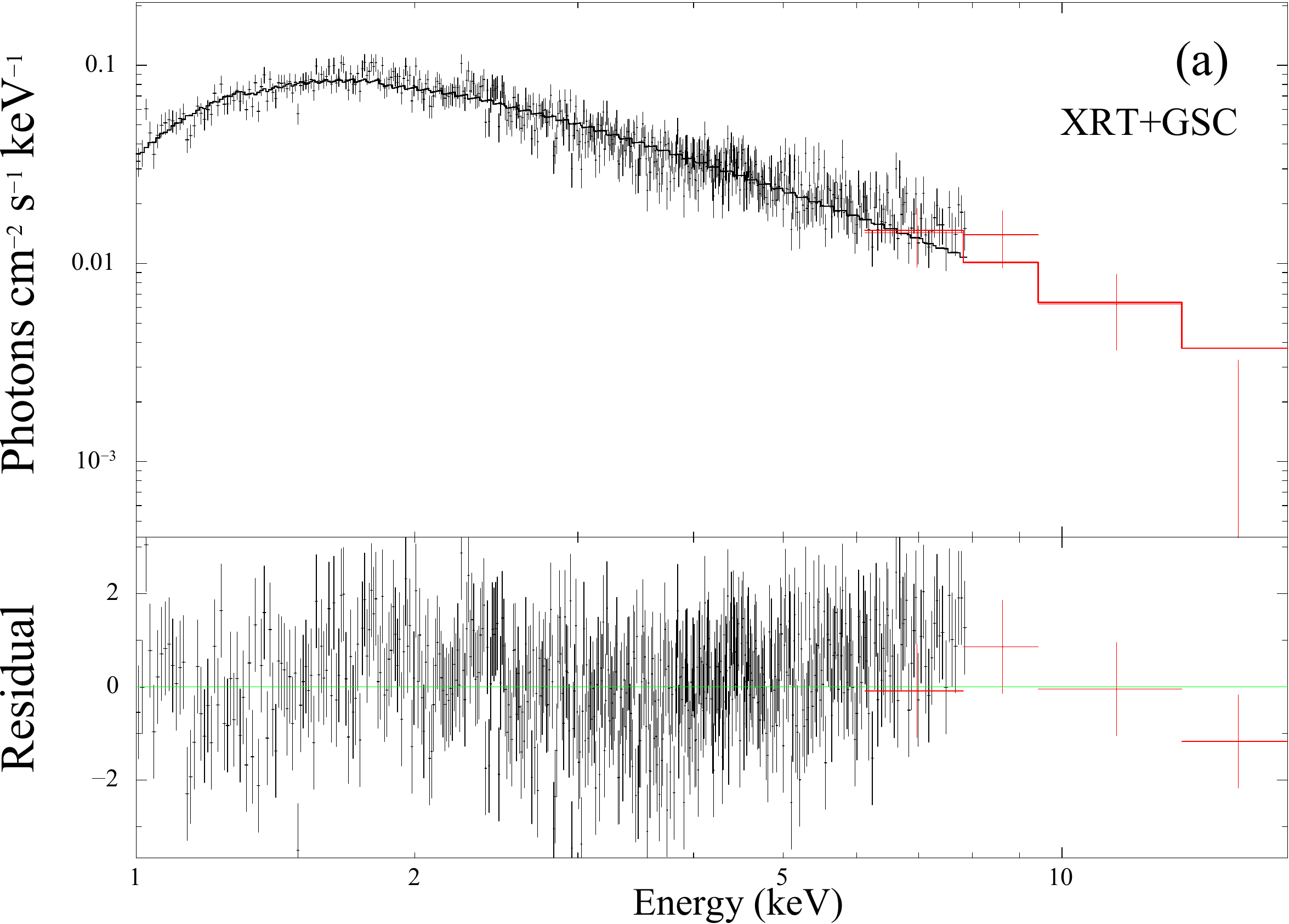} \hskip 0.6cm
\includegraphics[width=6.5truecm,angle=0]{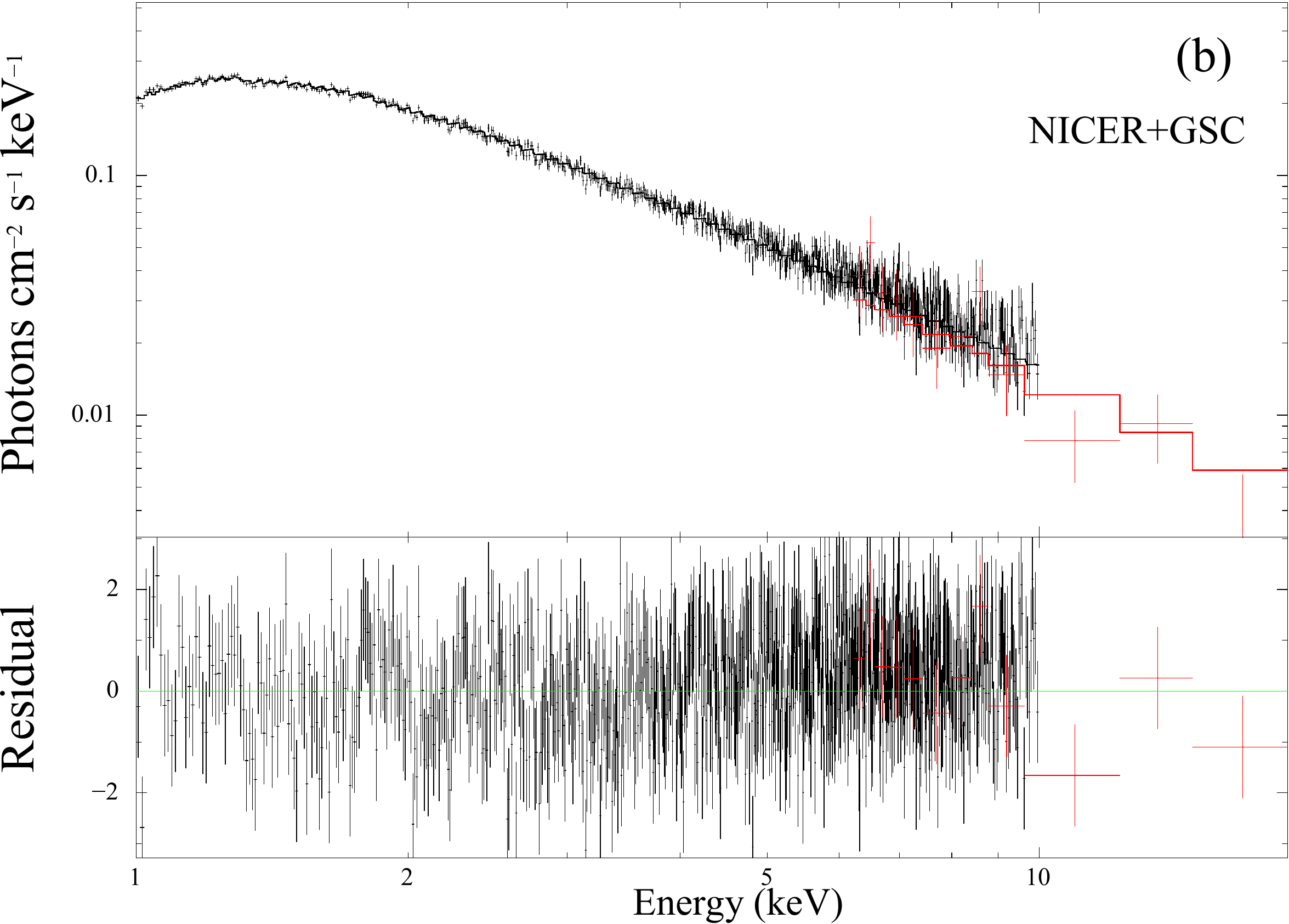}\\}
\vbox{
\includegraphics[width=6.5truecm,angle=0]{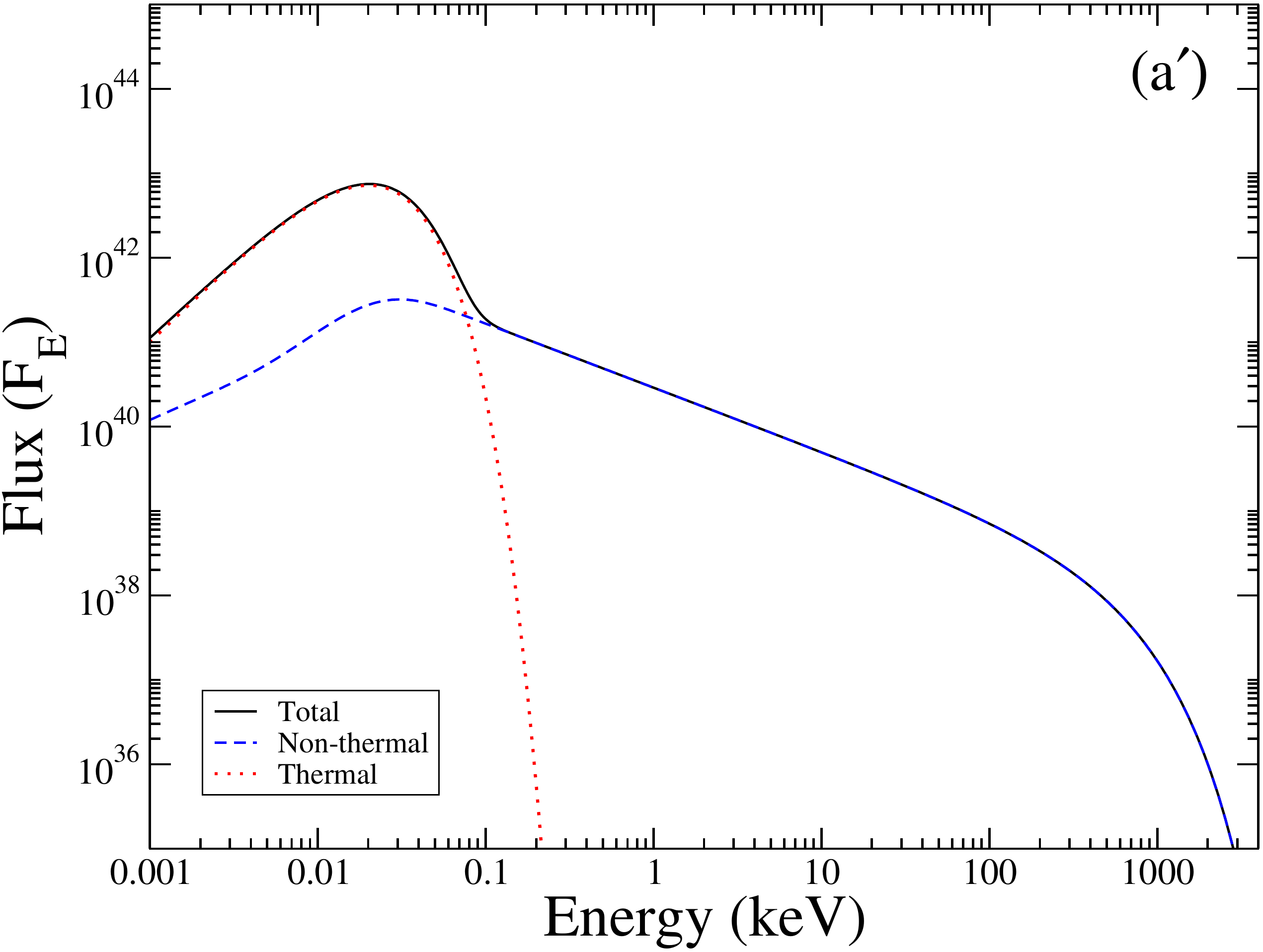} \hskip 0.6cm
\includegraphics[width=6.5truecm,angle=0]{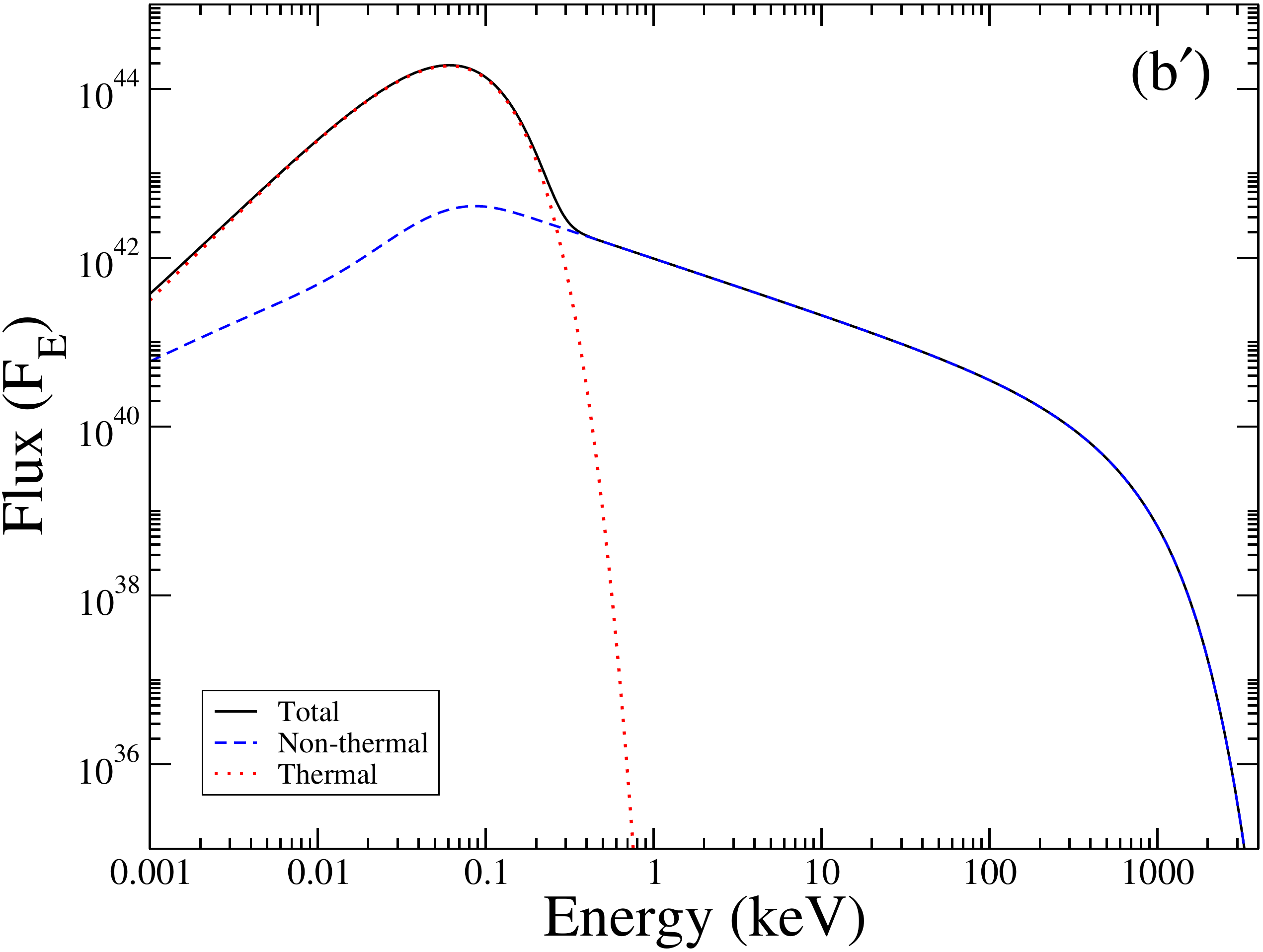}
}
\caption{TCAF model-fitted combined spectra of (\textbf{a}) Swift/XRT (OBSID 00011107047) with MAXI/GSC (MJD~=~58,690.23) in 1--20~keV in the top 
left panel, and (\textbf{b}) NICER (OBSID 2200530170) with MAXI/GSC (MJD~=~58,675.9) in 1--20~keV in the top right panel.
Bottom panels ({\boldmath{$a^\prime$}},{\boldmath{$b^\prime$}}) shows the unabsorbed TCAF model-generated spectra, which were used to fit the top panel spectra. Here, model flux ($F_E$) is plotted in 
units of photons~cm$^{-2}$~s$^{-1}$~keV$^{-1}$. Total flux is plotted with the solid curve. The~dotted red curve marks the thermal component through bremsstrahlung 
and Comptonization processes from the Keplerian disk of the pre-shock flow, and the dashed blue curve marks the non-thermal component through inverse Comptonization from the
hot Compton cloud or CENBOL.
\label{fig4}}
\end{figure}

Figure~\ref{fig4}a,b show TCAF model-fitted spectra using different spectral data. The~left panel shows the TCAF model-fitted
combined spectrum of Swift/XRT (OBSID 00011107047) with MAXI/GSC in 1--20~keV energy band during MJD~=~58,690.23, and the right
panel shows the TCAF model-fitted spectrum of the combined NICER (OBSID 2200530170) and simultaneous MAXI/GSC during
MJD~=~58,675.9 in 1--20~keV enery band. {The bottom panels of} Figure~\ref{fig4}{($a^\prime$,$b^\prime$) show two theoretical model unabsorbed raw spectra, which were used to fit observed spectra in the top panels. We can see the thermal and non-thermal components of the 
theoretical spectra separately in the plots. The~thermal part originates from the pre-shock flow via bremsstrahlung and Comptonization processes, 
while the non-thermal part of the spectra originates from the emitted photons from the hot Compton cloud or CENBOL via inverse Comptonization of 
the intercepted thermal photons} (see~\citep{CT95, DD15b} and references therein).

\begin{figure}[H]
\centering
\vbox{
\includegraphics[width=6.4truecm,angle=0]{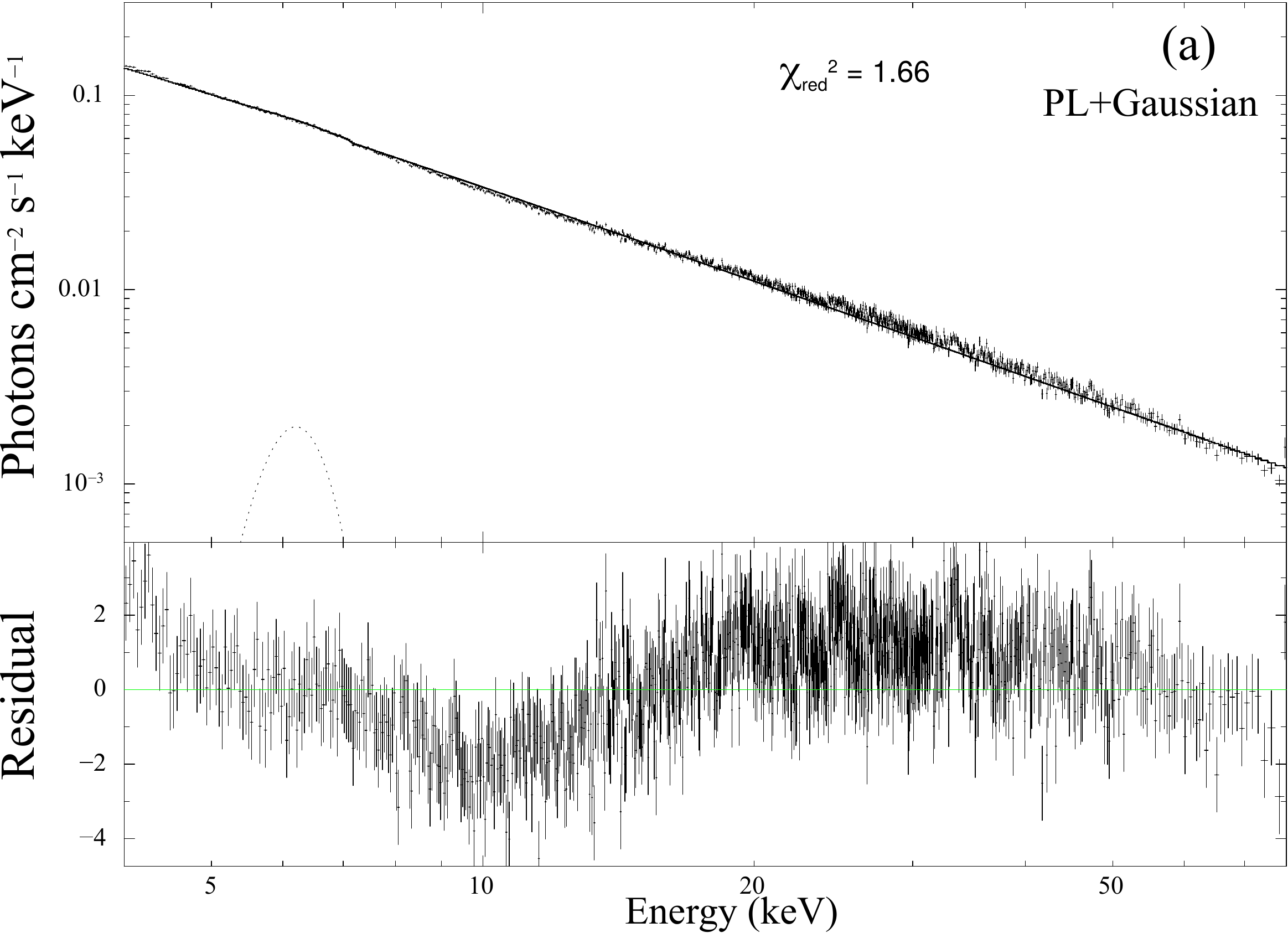}\hskip 0.6cm
\includegraphics[width=6.4truecm,angle=0]{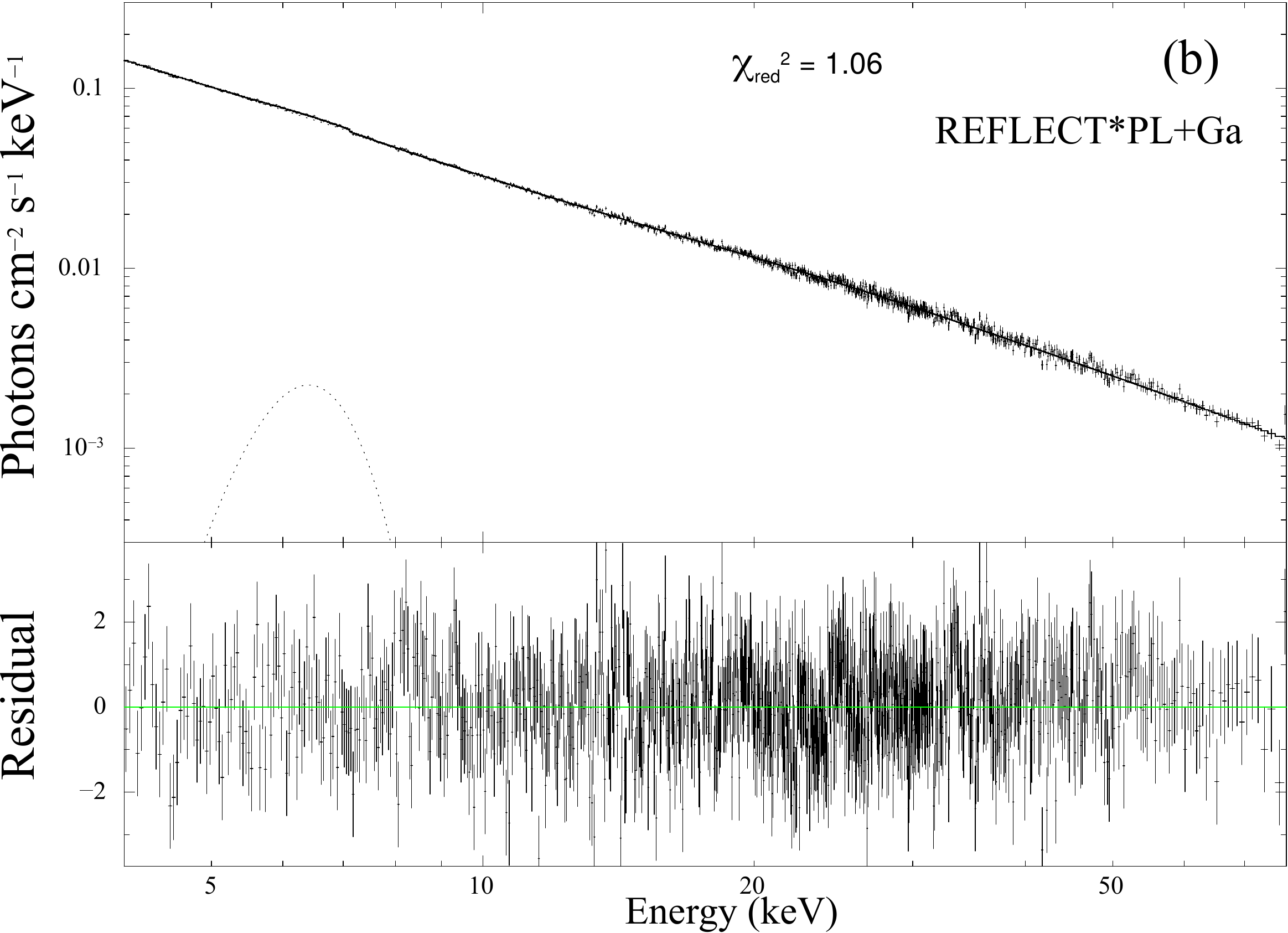}}
\vskip 0.3cm
\vbox{
\includegraphics[width=6.4truecm,angle=0]{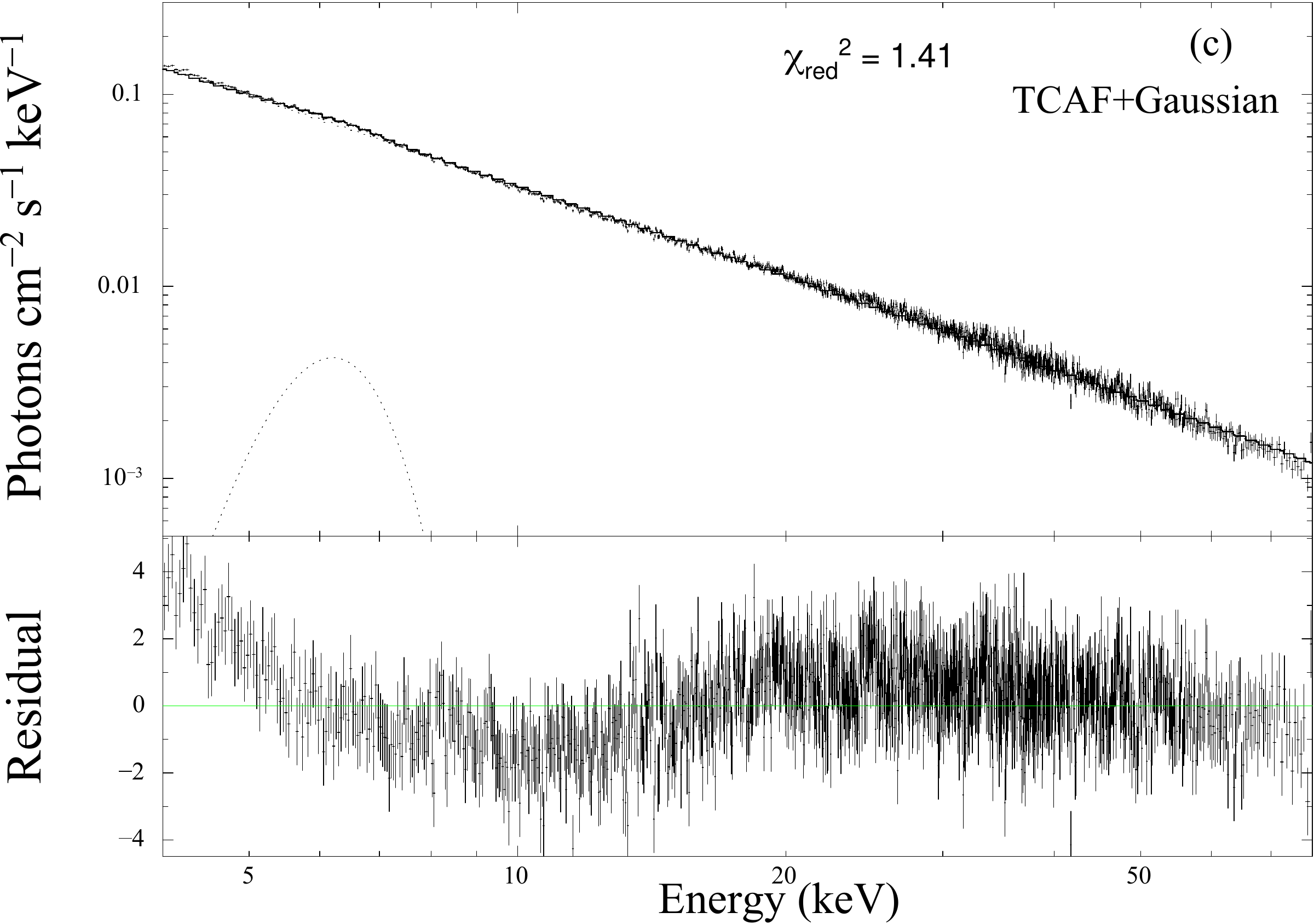}\hskip 0.6cm
\includegraphics[width=6.4truecm,angle=0]{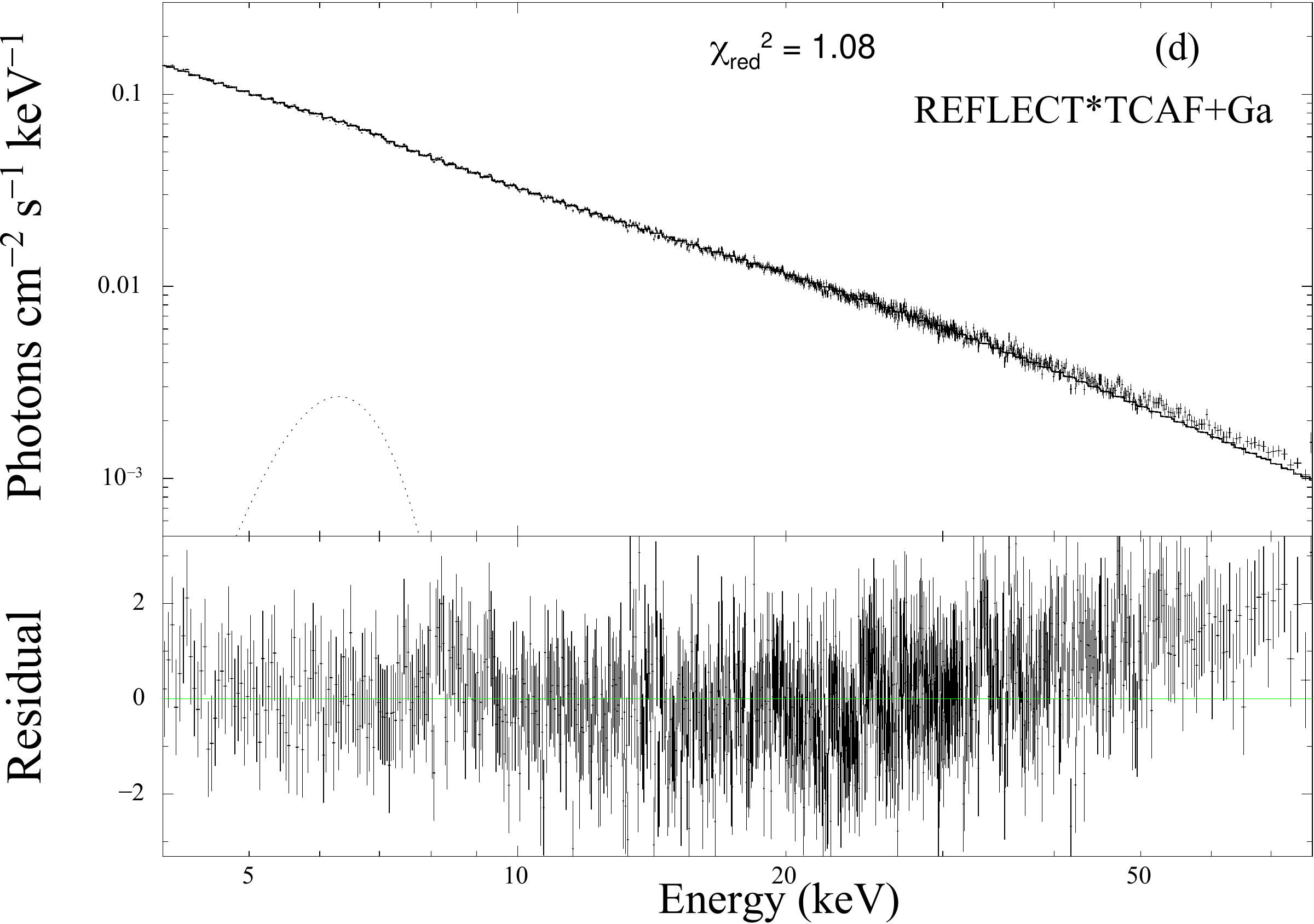}}
\caption{Fitted spectrum of NuSTAR of OBSID 80502304002 (MJD~=~58,655.60) in the energy range 4--78~keV with (\textbf{a}) PL+Gaussian model ,
(\textbf{b}) \textsc{reflect}*PL+Gaussian model, (\textbf{c}) TCAF+Gaussian model and (\textbf{d}) \textsc{reflect}*TCAF+Gaussian model.\label{fig5}}
\end{figure}

Figure~\ref{fig5} shows the fitted spectrum of the NuSTAR data (OBSID 80502304002) with different models. Panel (a) shows the 
PL+Gaussian model-fitted spectrum, which shows a signature of the reflection component in the disk $\sim$10~keV. 
Thus, we refitted the spectrum with \textsc{reflect}*powerlaw+Gaussian model (panel (b)). The~model-fitted reduced $\chi^2$ is improved 
from 1.66 to 1.06. This signifies the spectrum is fitted better with the \textsc{reflect}*powerlaw+Gaussian model. 
Panels (c) and (d) show the fitted spectrum of that particular observation with TCAF+Gaussian model and \textsc{reflect}*TCAF+Gaussian model.
In the last set, continuum due to primary emission is fitted with TCAF and \textsc{reflect}, where \textsc{reflect} is used to fit the reflection part 
(due to reprocessed photons) of the spectrum. It is clear from the figure that the spectrum is fitted better when we added 
the \textsc{reflect} model with the TCAF model to account for the reflection. The~reduced ${\chi}^2$ value decreased from~1.41 to 1.08.

\begin{figure}[H]
\centering
\includegraphics[width=6.6truecm,angle=0]{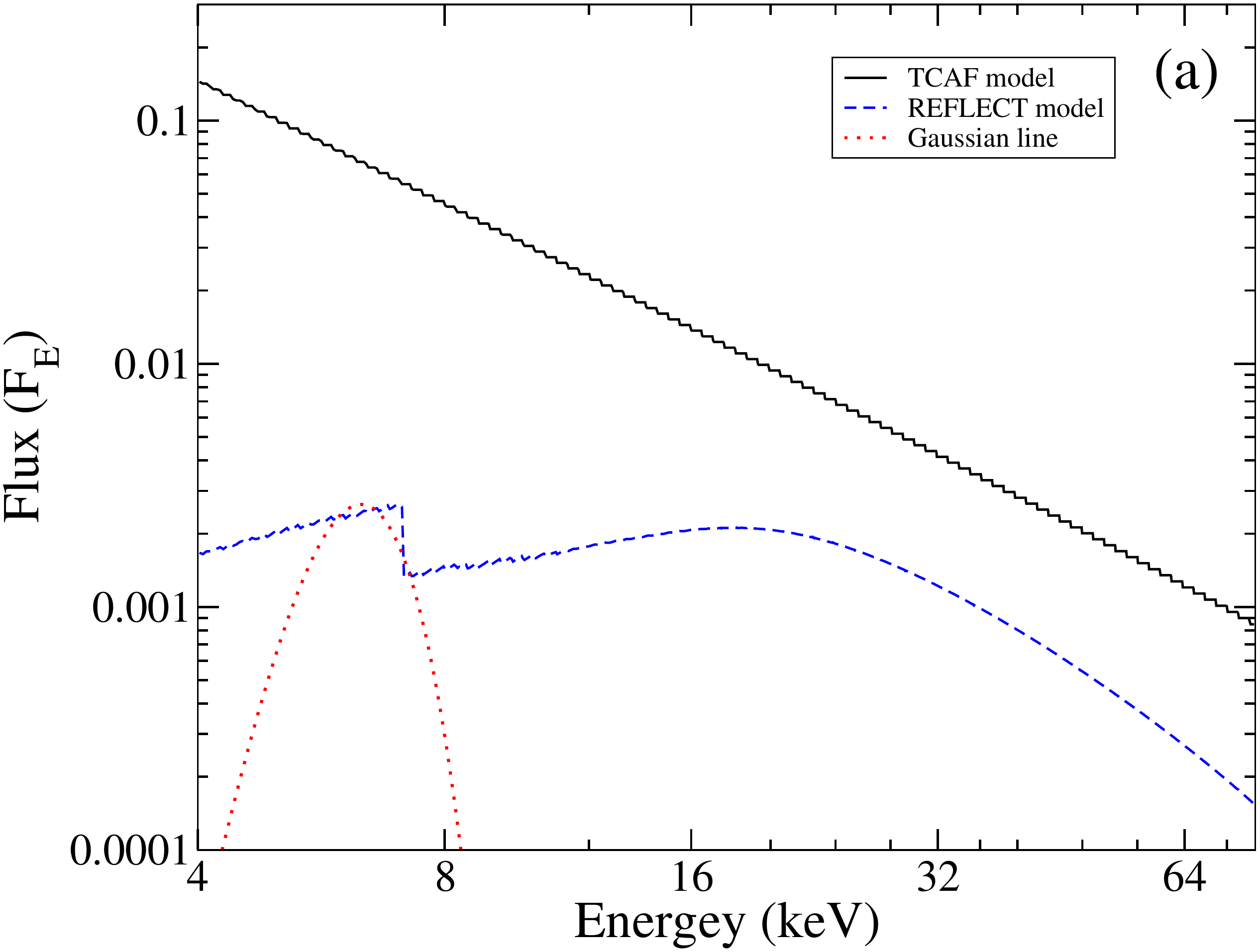} \hskip 0.3cm
\includegraphics[width=6.6truecm,angle=0]{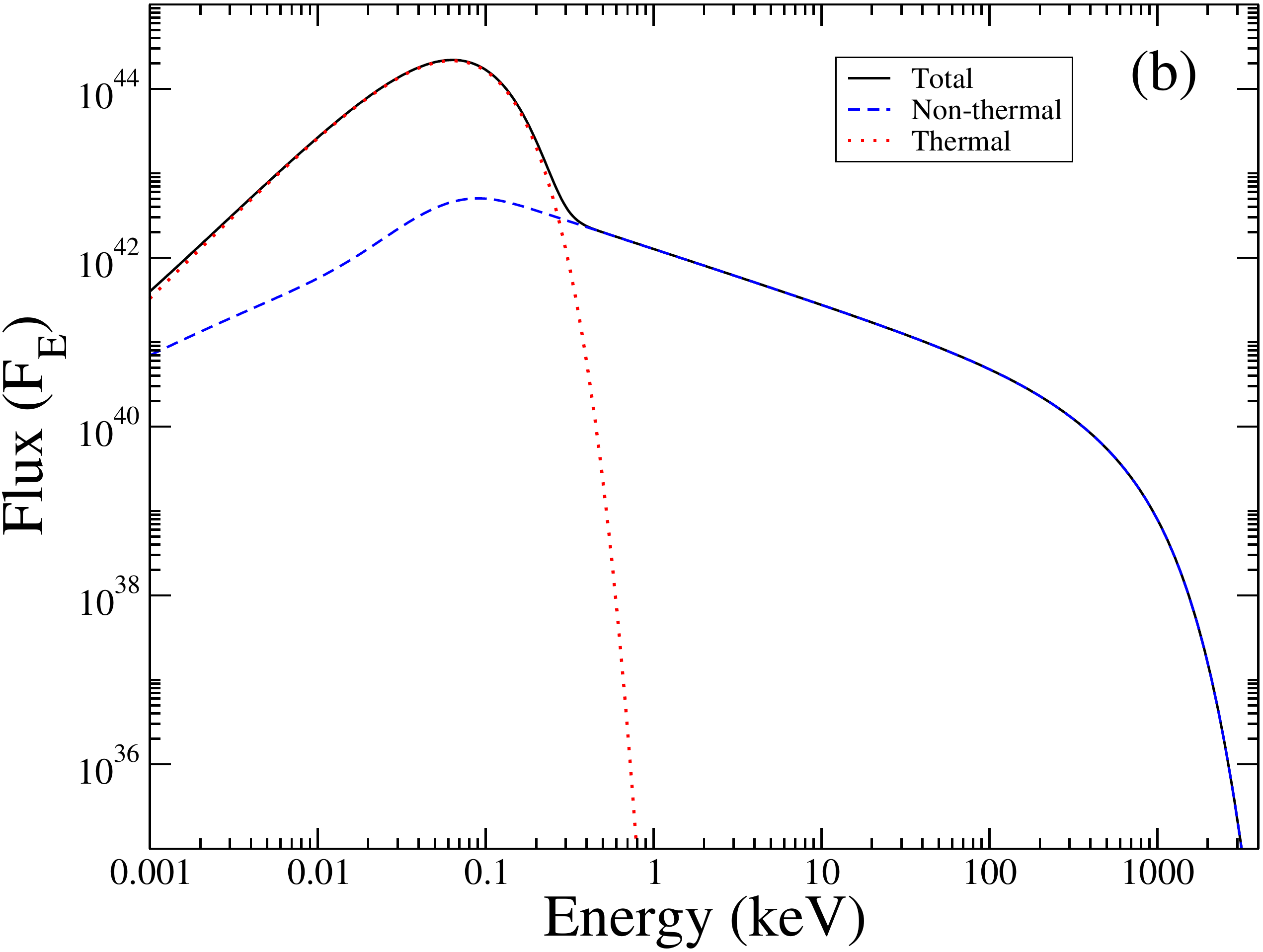}
	\caption{(\textbf{a}) Unabsorbed spectra of model components of the fitted spectrum shown in Figure~\ref{fig5}d for 
 the NuSTAR OBSID 80502304002 (MJD~=~58,655.60) from XSPEC, 
	where the contributions of the TCAF model (black solid curve), REFLECT model (blue dashed curve) and the Gaussian line (red dotted curve) are shown separately. 
	(\textbf{b}) Unabsorbed theoretical TCAF model raw spectrum with its two components, generated from the TCAF code using best-fitted TCAF model parameters of the same spectrum. 
	Flux ($F_E$) is plotted in photons~cm$^{-2}$~s$^{-1}$~keV$^{-1}$. \label{fig6}}
\end{figure}

Figure~\ref{fig6}a {shows the~unabsorbed model components of the fitted spectrum shown in} Figure~\ref{fig5}d {for the NuSTAR OBSID 
80502304002 (MJD~=~58,655.60) from XSPEC. Here, the contribution of the TCAF model, REFLECT model and the Gaussian line are shown separately. 
The theoretical TCAF model-generated unabsorbed raw spectrum with its two components (thermal and non-thermal) is shown in panel (b)} of Figure ~\ref{fig6}.

\begin{figure}[H]
\includegraphics[width=10.5 cm]{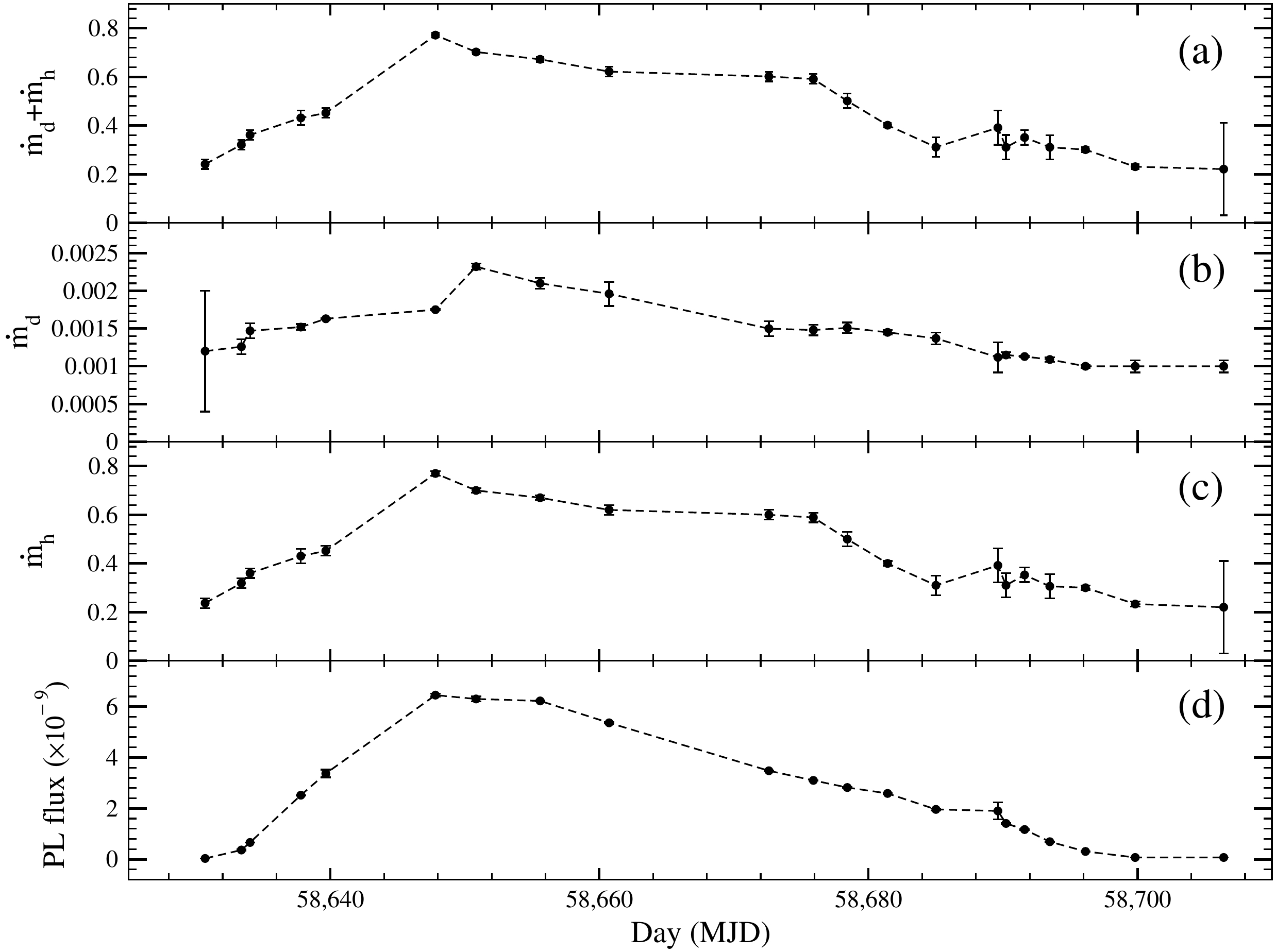}
	\caption{ Variation of TCAF model-fitted (\textbf{a}) total accretion rate ($\dot{m}_d$ + $\dot{m}_h$), (\textbf{b}) disk rate ($\dot{m}_d$) and (\textbf{c}) halo rate ($\dot{m}_h$). The~accretion rates are in units of Eddington rate ($\dot{M}_{Edd}$); (\textbf{d}) 2--10~keV powerlaw flux (obtained from powerlaw model) in $photons~cm^{-2} s^{-1}$.\label{fig7}}
\end{figure}

\begin{figure}[H]
\includegraphics[width=10.5 cm]{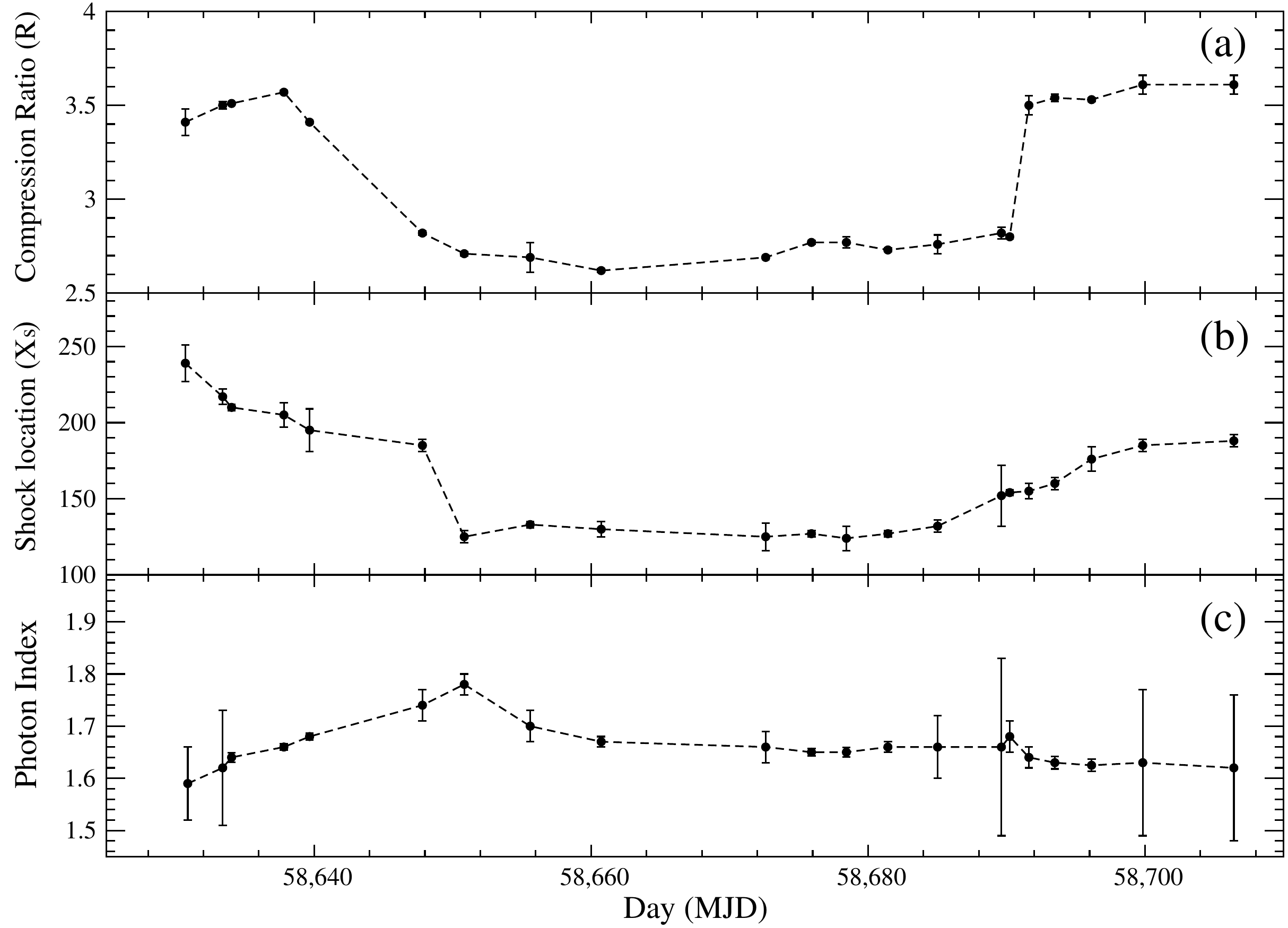}
\caption{ Variation of TCAF model-fitted (\textbf {a}) compression ratio (R), (\textbf{b}) shock location ($X_s$) and (\textbf{c})~powerlaw model-fitted photon index ($\Gamma$).\label{fig8}}
\end{figure}

Figures~\ref{fig7} and~\ref{fig8} show variation of the different model-fitted parameters. In~Figure~\ref{fig7}, we show the 
evolution of (a) the TCAF model-fitted total accretion rate ($\dot{m}_d$ + $\dot{m}_h$), (b) the Keplerian disk rate ($\dot{m}_d$) and 
(c) the sub-Keplerian halo rate ($\dot{m}_h$). Both the rates (disk rate and halo rate) are in units of Eddington rate ($\dot{M}_{Edd}$).
{Panel (d) of} Figure~\ref{fig7} {shows the variation of powerlaw flux.}
In Figure~\ref{fig8}, we show the variation of the TCAF model-fitted (a) shock compression ratio (R) and (b) the shock location ($X_s$) 
in ($r_s$). Panel (c) of Figure~\ref{fig8} shows the variation of the photon index.

The photon index of powerlaw ($\Gamma$) varied from 1.59--1.78 (Figure~\ref{fig8}c). On the first day of observation, the~value 
of $\Gamma$ was 1.59. Then, it gradually increased to a maximum value of 1.78 on MJD$\sim$58,650. After~that, $\Gamma$ started to decrease; it attained a minimum value of 1.63 on MJD$\sim$58,693 and remained there till the end of the~outburst. 

\textls[-15]{At the start of the outburst, the~halo accretion rate ($\dot{m}_h$) was$\sim$0.237 $\dot{M}_{Edd}$ (Figure~\ref{fig7}c). As~the 
outburst continued, the~$\dot{m}_h$ gradually increased and reached its maximum on MJD$\sim$58,648.} After~that, the~halo rate started to 
decrease and attained its minimum on MJD$\sim$58,685. The~$\dot{m}_h$ again increased briefly after MJD$\sim$58,685, and shortly after 
MJD$\sim$58,692, it declined again till the end of the outburst. At~the start of the outburst, the~disk rate ($\dot{m}_d$) was 
0.0012 $\dot{M}_{Edd}$ (very small compared to the halo rate). The~disk rate gradually increased and attained its maximum on MJD$\sim$58,651 
(3 days after the peak of the halo rate). Thereafter, the~disk rate started to decrease and attained the minimum value on MJD$\sim$58696
and remained at this value till the end of the observed period. Since the value of the disk rate was very small compared to the halo rate, 
the nature of the variation of total accretion rate ($\dot{m}_d$ + $\dot{m}_h$) is same as the halo rate. {The powerlaw continuum and its high-energy cutoff 
powerlaw models signify the non-thermal inverse-Comptonized high energy part of the spectra, whereas the $\dot{m}_h$ parameter of TCAF measures the 
rate of accretion of sub-Keplerian mass, which is responsible for forming the corona and inverse-Comptonization cloud. There is bound to be 
some correlation between $\dot{m}_h$ and powerlaw flux. Indeed, as~can be seen from Figure~\ref{fig7}d, the~powerlaw flux and the $\dot{m}_h$ 
shows similar variation during the course of the outburst. When $\dot{m}_h$ becomes high, it signifies the presence of large amounts of sub-Keplerian 
matter, which emits a large amount of inverse-Comptonized high-energy X-rays, which in turn increases the flux of the powerlaw part of the 
spectrum. When $\dot{m}_h$ becomes low, the~opposite phenomena happens, and powerlaw flux also reduces.}

\vspace{-10pt} 

In Figure~\ref{fig8}, it can be noticed that the shock compression ratio (R) is high, and the shock location ($X_s$) is far away from the BH 
at the start of the outburst ((a) and (b)). On~the first day (MJD$\sim$58,630) of the outburst, the shock ($X_s$) was located at a distance 
of $\sim$240$r_s$. The~$X_s$ started to decrease slowly with time and attained its minimum of $\sim$125$r_s$ on MJD$\sim$58,651. It remained 
between 125$r_s$--132$r_s$ till MJD$\sim$58,685, and after that, it started to increase gradually. On the last day of the outburst (MJD~=~58,706), 
the shock moved away at a distance of 188$r_s$. The~variation of the shock compression ratio (R), i.e.,~the ratio of post shock to pre-shock matter 
density, is shown in Figure~\ref{fig8}a. The~value of $R$ varied from 2.7 to 3.61. Initially, the value of R was around$\sim$3.5 at the 
start of the outburst. After~that, it decreased to a value 2.8 at MJD$\sim$58,647. The~value of R varied from 2.7--2.8 up to MJD $\sim$58,690,
and afterwards it started to increase. $R$ reached 3.61 on the last day of our~observation.

The spectral results depict the absence of the softer states during the outburst. The~photon index ($\Gamma$) never exceeded 1.8. 
Throughout the outburst, the~halo rate ($\dot{m}_h$) was dominant over the disk rate ($\dot{m}_d$). Although the~shock location moved slightly 
closer to the BH during the middle phase of the outburst, $X_s$ was never smaller than 125$r_s$. The~minimum value of the compression
ratio was 2.7 during the outburst. From~these results, we can conclude that BHC MAXI J1348-630 went through the hard state (HS) only during the 
second outburst of 2019 (May 2019 to August 2019). 

\begin{figure}[H]
\vbox{
\includegraphics[width=6.6truecm,angle=0]{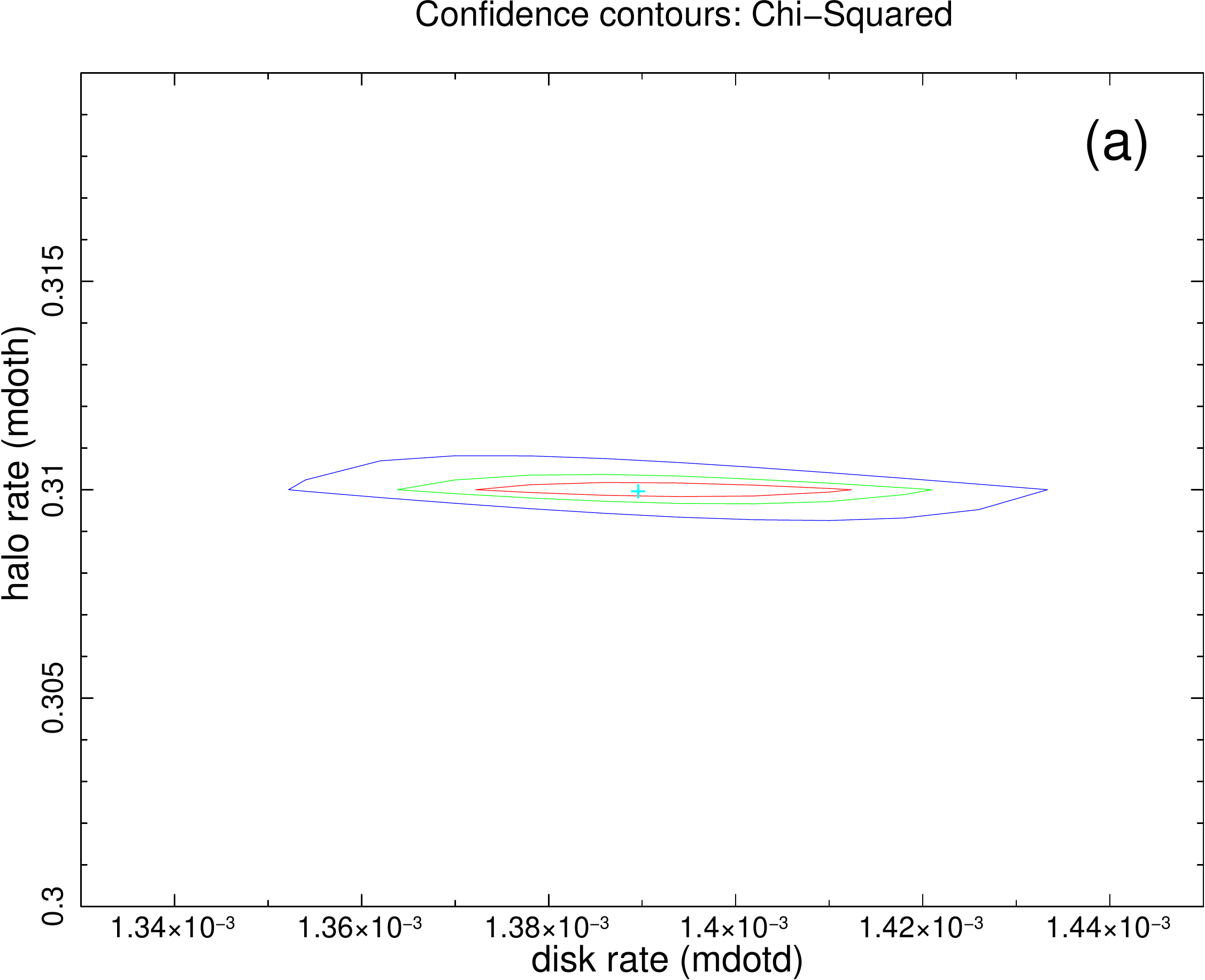}\hskip 0.4cm
\includegraphics[width=6.6truecm,angle=0]{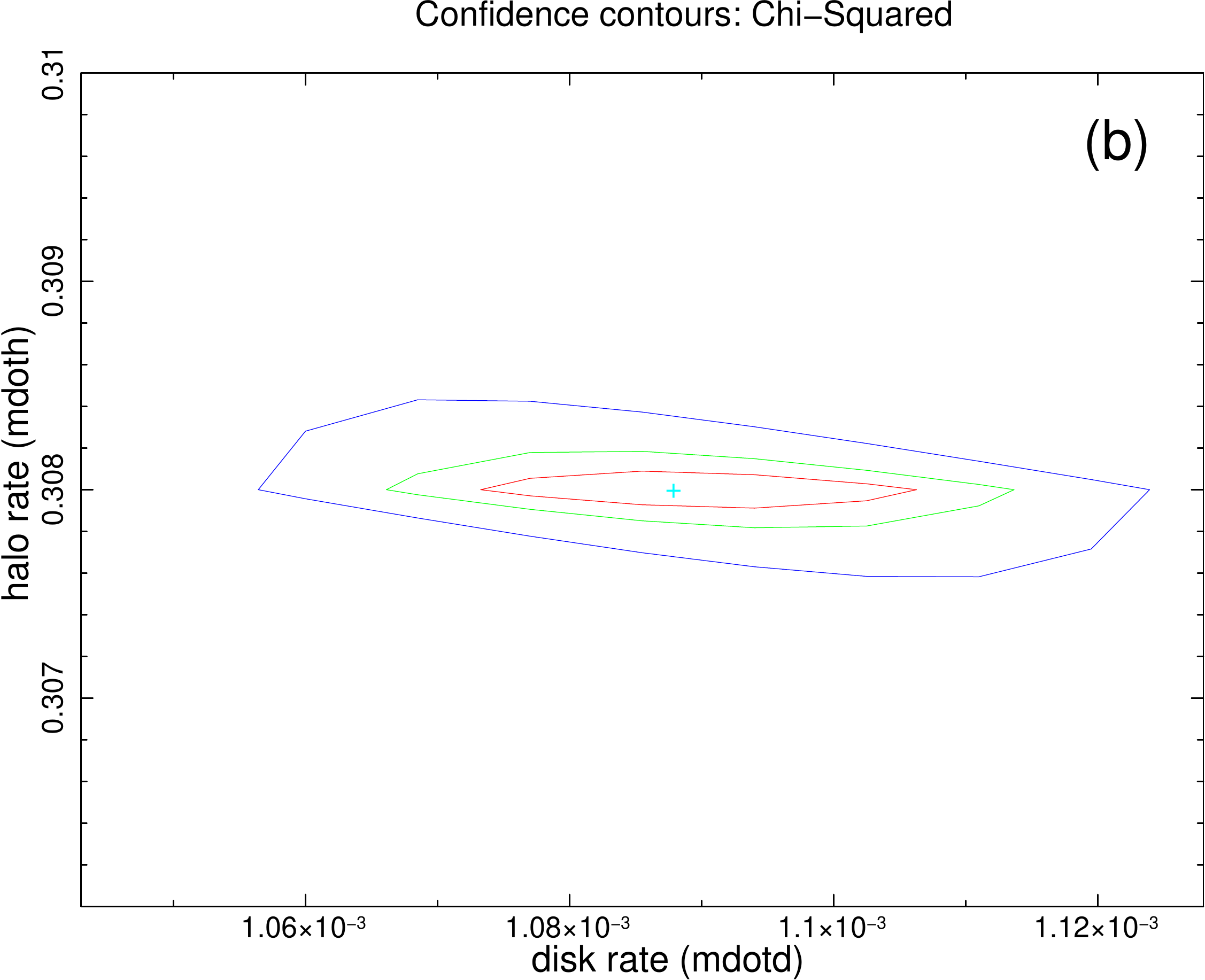}}
\caption{Confidence contours of disk rate ($\dot{m}_d$) vs. halo rate ($\dot{m}_h$) for two datasets: (\textbf{a}) Swift/XRT Obs. ID~=~00011107045 (MJD~=~58,685) in the left panel and (\textbf{b}) the combined data of NICER Obs. ID~=~2200530187 with MAXI/GSC (MJD~=~58,693) in the right~panel.\label{fig9}}
\end{figure}

We kept hydrogen column 
density ($n_H$) free during our analysis. The~$n_H$ varied within a range of 0.50 $\times$ 10$^{22}$--0.93 $\times$ $10^{22}$ during the observation period.
The 2D contour plots of disk rate ($\dot{m}_d$) vs. halo rate ($\dot{m}_h$) for Swift Observation ID 00011107045 (MJD~=~58,685) and for the combined data of
NICER (Observation ID 2200530187) and MAXI/GSC (MJD~=~58,693) are shown in Figure~\ref{fig9}.

For NuSTAR spectra,
we found the presence of weak reflection, with the reflection fraction ($R_{refl}$) in the range of 0.23--0.32. 
The reflection fraction was found to be between 0.10--0.15 by fitting the NuSTAR observations using the relxillCp model
\citep{Jia2022}. The~far away location of the Keplerian disk and the lower value of the accretion rate may be the reasons behind this
weak reflection. The~cos(incl) angle varied between 0.70--0.86, which depicts the inclination angle as $\sim$30--46 deg.
Our estimation of inclination angle is consistent with a previous report of inclination angle of $i$~=~30--40 degrees~\citep{Chakraborty2021}.

\vspace{-10pt}

\vspace{-10pt}

\subsubsection{Viscous Time~Scale} 
The time taken for high-viscosity matter to travel from the pile up radius to the BH is known as the viscous time scale~(\citep{AJ16} and references 
therein). The~low-viscosity sub-Keplerian matter or halo moves roughly in freefall time scale, whereas the high-viscosity Keplerian matter moves slowly 
in viscous time scale. As~the halo moves faster than the Keplerian matter, it reaches the BH earlier than the Keplerian matter in the rising phase 
when an outburst occurs. As~a result, the halo rate peaks before the disk rate. Thus, the viscous time scale can be calculated by determining 
the difference between the peaks of the two accretion rates (for more details, see~\citep{AJ16}). In~the present outburst, it can be seen that the 
halo rate became maximum on MJD~=~58,648, whereas the~disk rate attained its maximum value 3 days later (MJD~=~58,651) (Figure~\ref{fig7}b,c). Considering this, we can conclude that the viscous time scale for this outburst is roughly 3~days.

\section{Discussion and~Conclusions}
\label{sect:discussion}
The galactic BHC MAXI~J1348-630 was discovered on 2019 January 26 by MAXI/GSC. It has shown two large outbursts (from MJD~=~58,509 to MJD~=~58,610 and from
MJD~=~58,630 to MJD~=~58,700) and six mini outbursts since its discovery~\citep{Negoro2020, Baglio2020, Carotenuto2020}.  
In Jana~et~al. (2020)~\citep{AJ20}, the~timing and spectral properties of the first outburst (MJD~=~58,509 to MJD~=~58,610) were studied in detail with the physical 
TCAF model. In~this paper, we studied the spectral and timing properties of the source during its second outburst (MJD~=~58,630 to MJD~=~58,700) in 2019. 
We performed spectral analysis of MAXI~J1348-630 using Swift/XRT(1--8~keV), NICER (1--10~keV), MAXI/GSC (6--20~keV) 
and NuSTAR (4--78~keV) archival data. For~the spectral analysis, we combined MAXI/GSC with Swift/XRT and NICER data and analyzed the 1--20~keV 
energy band. Swift/XRT+MAXI/GSC and NICER+MAXI/GSC were first fitted with the phenomenological powerlaw (PL) model and then with the 
physical TCAF model. We studied the NuSTAR data separately. NuSTAR data was fitted with \textsc{reflect}*powerlaw+Gaussian model and 
\textsc{reflect}*TCAF+Gaussian~model.  

From the spectral analysis, we classified the spectral nature of the source during the outburst. From~the low value of the photon index ($\Gamma<$1.8),
we can roughly say that the source remained in the harder states. In~softer states, the powerlaw photon index is much 
higher ($\gtrsim2$). Further, from~the TCAF analysis, we observed that throughout the outburst, the disk rate was very low 
(in the order of $\sim$$10^{-3}$) compared to the halo rate. Moreover, the shock did not come close to the BH and was found to be at $\gtrsim$125$r_s$ 
with high $R$ ($\gtrsim$2.7). In~softer states, one would expect a weak shock to be located close to the BH. From~these spectral parameters, we conclude that the source remained in the hard state (HS) throughout the outburst. In~all three observations of
NuSTAR spectra, we found the presence of weak reflection. The~distant location of the Keplerian disk may be the reason for this weak reflection.
From the \textsc{reflect} model, we found that the inclination angle of the source varied from 30$^\circ$--46$^\circ$.

In the harder states of BHs, low-frequency quasi-periodic oscillations are very commonly observed. We performed timing analysis using 
LAXPC (3--80~keV) and NICER (1--10~keV) 0.01 s time binned light curves. We found QPOs in two successive dates, 2019 June 14 and 2019 
June 15, with QPO frequencies of 0.96~Hz and 0.95~Hz, respectively using the LAXPC data. We did not observe any
prominent QPO nature in the NICER data. {From 2019 June 14 data, we also found a 
sub-harmonic of the primary QPO at 0.5~Hz. The~same nature of QPOs have also been found in BHs XRBs XTE J1550-564 and XTE J1859+226}~\citep{Casella2004, Remillard2002a}.
{Neither of the two observed QPOs (2019 June 14 and 2019 June 15) fit the mold, i.e.,~generalized types (A, B or C) of QPOs mentioned in} 
\citep{Casella2005, Motta2016, Remillard2002b}. {Thus, we are unable to classify the types of the observed QPOs. Shang~et~al. 2019}~\citep{Shang2019} {also found an unknown type of LFQPO around $\sim$0.41~Hz for the BHC MAXI J1535-571.}

We performed a detailed spectral and temporal analysis of the second outburst of the BHC MAXI J1348-630 in 
this paper. Although~a detailed study of the first outburst has been completed by many authors, the~second outburst was 
less-studied. The~earlier (first) outburst of MAXI J1348-630 was a complete or normal type outburst, during which
all spectral states were found to form a hysteresis loop in the following sequence: HS $\rightarrow$ HIMS $\rightarrow$
SIMS $\rightarrow$ SS $\rightarrow$ SIMS $\rightarrow$ HIMS $\rightarrow$ HS (Jana~et~al. 2020). However, here the source was 
found only in HS. No evolution of the spectral states was observed. Thus, we termed this outburst as a failed outburst. 
Although there are various papers in the literature, no detailed study covering the entire outburst hsa been performed for this particular outburst. We found QPOs in two orbits (20064 and 20073) of AstroSat data (observation ID= T03 120T01 9000002990), 
while no QPO was found in NICER observations. We also found the presence of a weak reflection component in the NuSTAR spectra. 
We found that only the physical TCAF model was not able to fit these spectra. This was also an important finding from our analysis.
We also predicted the value of the inclination angle of the source to be in the range of 30$^\circ$--46$^\circ$, which is consistent
with the previous report by Chakraborty~et~al. (2021)~\citep{Chakraborty2021}.

In our recent studies on understanding the triggering mechanism of an outburst of transient BHCs, we predicted that 
the accreted matter from the companion star accumulates at the pile-up radius ($X_p$) during the quiescence phase 
prior to the start of an outburst~\citep{SKC19}. A~good linear relation between outbursts with quiescent 
(or accumulation) periods was established while studying recurring transient BH X-ray binaries: H 1743-322 
\citep{SKC19}, GX 339-4~\citep{RB21}. As~the amount of accumulated matter increases at $X_p$,
thermal pressure rises, which in turn increases the turbulence and creates instability in the disk. Due to this, viscosity
rises, and~when it exceeds a certain critical value at this temporary reservoir, matter starts to accrete, and an
outburst is triggered~\citep{Ebisawa96}. When viscosity falls below a critical value, the~accretion stops. All the
matter that had accumulated before an outburst may not have been cleared during the outburst. Hence, the matter remains stuck at the $X_p$.
The $X_p$ may move closer to the BH during the outburst. The~leftover matter gets combined with freshly supplied matter
from the companion. Once enough matter has accumulated, instability may trigger another outburst when the viscosity rises
above the critical value again. The~smaller the $X_p$, the~shorter the quiescence phase. This is because a smaller $X_p$ requires a lower viscosity to trigger the outburst, and so less mass accumulation is sufficient to trigger the outburst~\citep{SKC19, RB21}. The~quiescence period
between the first and second outbursts of BHC MAXI~J1348-630 was comparatively small at $\sim$20~days. The~second outburst was a 'failed outburst' and continued for almost two and half months. Thus, the scenario here may be that
all the matter that had accumulated prior to the first outburst was not cleared during the first outburst. The~leftover
matter combined with freshly supplied matter and triggered the second outburst after a very small quiescent~period.

\vspace{6pt}

\authorcontributions{
Conceptualization, R.B., D.D., A.J. and K.C.; data curation, R.B. and S.K.N.; formal analysis, R.B., D.D. and A.J.; methodology, R.B., D.D. and A.J.; software, D.D.; supervision, D.D.; writing---original draft, R.B.; writing---review and editing, R.B., D.D., A.J., K.C. and S.K.N. All authors have read and agreed to the published version of the~manuscript.}

\funding{
This research received no external funding. The APC is not funded. The agencies, which are mentioned in the acknowledgment section, fund the salaries and fellowships of the authors. 
}

\dataavailability{
This research used data and/or software provided by the High Energy Astrophysics Science Archive Research Center (HEASARC),
which is a service of the Astrophysics Science Division at NASA/GSFC. 
This work used Swift/XRT data supplied by the UK Swift Science Data Centre at the University of Leicester; MAXI/GSC data provided 
by RIKEN, JAXA and~the MAXI team; NICER data archived by NASA/GSFC; NuSTAR data by NASA/GSFC; and AstroSat/LAXPC data obtained from the data archive of the Indian Space 
Science Data Centre (ISSDC). We acknowledge the strong support from the Indian Space Research Organization (ISRO) for the successful realization and operation of the AstroSat mission. The~authors also acknowledge the AstroSat team for the distribution. LaxpcSoft 
software was used for the analysis.}

\acknowledgments{R.B. acknowledges support from the CSIR-UGC fellowship (June-2018, 527223). D.D. acknowledges support
from the DST/GITA-sponsored India--Taiwan collaborative project (GITA/ DST/TWN/P-76/2017). Research of D.D. is supported in part by the Higher
Education Dept. of the Govt. of West Bengal, India. K.C. acknowledges support from the DST/INSPIRE (IF170233) fellowship. A. J. acknowledges
the support of a grant from the Ministry of Science and Technology of Taiwan, with grant numbers MOST 110-2811-M-007-500 and 
MOST 111-2811-M-007-002. S.K.N. and D.D. acknowledge partial support from the ISRO-sponsored RESPOND project (ISRO/RES/2/418/17-18)~fund.
S.K.N. acknowledges the SVMCM scholarship, Government of West Bengal.} 

\conflictsofinterest{The authors declare no conflict of~interest.} 







\appendixtitles{no} 
\appendixstart
\appendix
\section[\appendixname~\thesection]{}
\begin{table}[H]
\caption{Model-fitted
parameters for Swift/XRT, combined NICER+MAXI/GSC and Swift/XRT+MAXI/GSC spectra (taking mass($M_{BH}$) as a free parameter for TCAF model). \label{ta1}}
\begin{adjustwidth}{-\extralength}{0cm}
\addtolength{\tabcolsep}{+1.0pt}
		\newcolumntype{C}{>{\centering\arraybackslash}X}
		\begin{tabularx}{\fulllength}{CCCCCCCCCCC}
\toprule
\boldmath{\textbf{ID}$^{[1]}$} & \boldmath{$n_H^{[2]}$} & \boldmath{$\Gamma^{[3]}$} & \boldmath{$flux^{[3]}$} & \boldmath{${\chi}^2$/dof$^{[5]}$} & \boldmath{${\dot m}_d$$^{[4]}$} & \boldmath{${\dot m}_h$$^{[4]}$} & \boldmath{R$^{[4]}$} & \boldmath{$X_s$$^{[4]}$} & \boldmath{$M_{BH}$$^{[4]}$} & \boldmath{${\chi}^2$\textbf{/dof}$^{[5]}$} \\
 \textbf{(1)}        & \textbf{(2)}      & \textbf{(3)}     & \textbf{(4)}           & \textbf{(5)}           & \textbf{(6)}     & \textbf{(7)}              & \textbf{(8)}            & \textbf{(9)}           & \textbf{(10)} & \textbf{(11)} \\
\midrule
X1  & $0.64^{\pm0.20}$ & $1.59^{\pm0.07}$ & $0.36^{\pm0.04}$ &22/27   & $1.20^{\pm0.08}$ & $0.24^{\pm0.19}$ & $3.41^{\pm0.04}$ & $239^{\pm14}$  & $10.1^{\pm0.53}$ & 19/23\\
X2  & $0.78^{\pm0.14}$ & $1.62^{\pm0.11}$ & $3.67^{\pm0.25}$ &199/197 & $1.31^{\pm0.10}$ & $0.32^{\pm0.02}$ & $3.50^{\pm0.02}$ & $219^{\pm5}$  & $7.9^{\pm0.21}$ & 197/193\\
NI1 & $0.66^{\pm0.05}$ & $1.64^{\pm0.01}$ & $6.64^{\pm0.03}$ &764/709 & $1.37^{\pm0.03}$ & $0.34^{\pm0.02}$ & $3.60^{\pm0.002}$ & $210^{\pm1}$ & $7.9^{\pm0.05}$ & 776/705\\
NI2 & $0.51^{\pm0.03}$ & $1.66^{\pm0.01}$ & $25.2^{\pm0.10}$ &924/839 & $1.50^{\pm0.04}$ & $0.38^{\pm0.02}$ & $3.57^{\pm0.003}$ & $205^{\pm2}$ & $7.9^{\pm0.03}$ & 1011/835\\
X3  & $0.65^{\pm0.09}$ & $1.68^{\pm0.06}$ & $33.7^{\pm1.50}$ &721/622 & $1.63^{\pm0.30}$ & $0.45^{\pm0.03}$ & $3.41^{\pm0.01}$ & $195^{\pm13}$  & $8.1^{\pm0.08}$ & 724/618\\
X4  & $0.63^{\pm0.06}$ & $1.74^{\pm0.03}$ & $64.5^{\pm0.50}$ &959/686 & $1.79^{\pm0.08}$ & $0.78^{\pm0.01}$ & $2.80^{\pm0.06}$ & $180^{\pm14}$ & $10.3^{\pm0.08}$ & 931/682\\
X5  & $0.53^{\pm0.13}$ & $1.78^{\pm0.02}$ & $63.0^{\pm1.00}$ &707/627 & $2.32^{\pm0.04}$ & $0.67^{\pm0.05}$ & $2.70^{\pm0.04}$ & $125^{\pm1}$  & $10.2^{\pm0.16}$ & 723/623\\
NI3 & $0.51^{\pm0.02}$ & $1.65^{\pm0.01}$ & $31.0^{\pm0.10}$ &788/779 & $1.48^{\pm0.07}$ & $0.59^{\pm0.03}$ & $2.77^{\pm0.01}$ & $127^{\pm2}$  & $8.0^{\pm0.02}$ & 811/775\\
NI4 & $0.52^{\pm0.02}$ & $1.65^{\pm0.01}$ & $28.2^{\pm0.20}$ &1037/872& $1.50^{\pm0.07}$ & $0.50^{\pm0.03}$ & $2.77^{\pm0.03}$ & $124^{\pm7}$  & $8.5^{\pm0.02}$ & 1046/868\\
NI5 & $0.55^{\pm0.03}$ & $1.66^{\pm0.01}$ & $25.9^{\pm0.20}$ &979/841 & $1.45^{\pm0.04}$ & $0.40^{\pm0.01}$ & $2.73^{\pm0.01}$ & $127^{\pm1}$  & $7.9^{\pm0.03}$ & 964/837\\
X6  & $0.79^{\pm0.20}$ & $1.66^{\pm0.06}$ & $19.6^{\pm0.40}$ &561/522 & $1.39^{\pm0.02}$ & $0.31^{\pm0.04}$ & $2.76^{\pm0.01}$ & $132^{\pm1}$  & $7.9^{\pm0.11}$ & 569/518\\
X7  & $0.50^{\pm0.29}$ & $1.66^{\pm0.17}$ & $19.0^{\pm3.30}$ &57/48   & $1.12^{\pm0.20}$ & $0.39^{\pm0.07}$ & $2.82^{\pm0.03}$ & $152^{\pm21}$ & $10.2^{\pm0.87}$ & 54/44\\
X8  & $0.93^{\pm0.25}$ & $1.68^{\pm0.03}$ & $14.1^{\pm0.20}$ &596/489 & $1.15^{\pm0.04}$ & $0.30^{\pm0.07}$ & $2.80^{\pm0.01}$ & $153^{\pm12}$ & $7.9^{\pm0.07}$ & 596/485\\
NI6 & $0.61^{\pm0.01}$ & $1.64^{\pm0.02}$ & $11.7^{\pm0.20}$ &771/801 & $1.13^{\pm0.01}$ & $0.35^{\pm0.09}$ & $3.50^{\pm0.01}$ & $155^{\pm18}$  & $8.0^{\pm0.02}$ & 894/797\\
NI7 & $0.60^{\pm0.02}$ & $1.63^{\pm0.01}$ & $6.91^{\pm0.03}$ &612/637 & $1.09^{\pm0.40}$ & $0.31^{\pm0.01}$ & $3.54^{\pm0.02}$ & $160^{\pm19}$  & $9.4^{\pm0.03}$ & 639/633\\
NI8 & $0.63^{\pm0.04}$ & $1.62^{\pm0.01}$ & $3.11^{\pm0.05}$ &469/517 & $1.00^{\pm0.03}$ & $0.30^{\pm0.01}$ & $3.53^{\pm0.17}$ & $176^{\pm6}$  & $8.2^{\pm0.05}$ & 468/513\\
X9  & $0.83^{\pm0.13}$ & $1.63^{\pm0.14}$ & $0.71^{\pm0.04}$ &55/50   & $1.00^{\pm0.09}$ & $0.23^{\pm0.01}$ & $3.60^{\pm0.05}$ & $187^{\pm3}$  & $8.1^{\pm0.33}$ & 56/46\\
X10 & $0.84^{\pm0.04}$ & $1.62^{\pm0.14}$ & $0.72^{\pm0.04}$ &64/52   & $1.00^{\pm0.08}$ & $0.22^{\pm0.19}$ & $3.61^{\pm0.05}$ & $188^{\pm4}$  & $7.9^{\pm0.44}$ & 64/48\\
\bottomrule
\end{tabularx}
\end{adjustwidth}
\noindent{\footnotesize
{$^{[1]}$ ID of the observed dates as mentioned in Table~\ref{tab1} (Col. 1).}
{$^{[2]}$ Model-fitted value of hydrogen column density ($n_H$) in $10^{22}$ atoms per cm$^{-2}$ (Col. 2).}
{$^{[3]}$ PL model-fitted photon index ($\Gamma$); Col. 3.}
{$^{[3]}$ PL model-fitted flux; Col. 4 in $10^{-10}$ order}
{$^{[4]}$ TCAF model-fitted parameters: disk rate (${\dot m}_d$ in Eddington rate ${\dot M}_{Edd}$) in $10^{-3}$ order,} 
{halo rate (${\dot m}_h$ in ${\dot M}_{Edd}$), compression ratio ($R$), shock location ($X_s$ in Schwarzschild radius $r_s$)}
{and mass of the black hole ($M_{BH}$ in solar mass $M_{\odot}$); Cols. 6--10, respectively.}
{$^{[5]}$ PL and TCAF model-fitted ${\chi}^2_{red}$ values; Cols. 5 and 11, respectively, as}
{${\chi}^2/dof$, where `dof' represents degrees of freedom.}
{Note: We present average values of 90\% confidence $\pm$ parameter error values, which are}
{obtained using `err' task in XSPEC.}
}
\end{table}
\unskip

\begin{table}[H]
\tablesize{\fontsize{6.5}{6.5}\selectfont}
\caption{Fitted
 parameters for NuSTAR data with TCAF/\textsc{reflect}*TCAF model along with a Gaussian line (taking mass($M_{BH}$) as a free parameter). \label{ta2}}
\begin{adjustwidth}{-\extralength}{0cm}
		\newcolumntype{C}{>{\centering\arraybackslash}X}
		\begin{tabularx}{\fulllength}{ccCCCCCCCCCCCC}
\toprule
&\boldmath{\textbf{ID}$^{[1]}$}  & \boldmath{$n_H$$^{[2]}$} & \boldmath{$rel_{refl}$$^{[3]}$} & \boldmath{$cosIncl$$^{[3]}$} & \boldmath{${\dot m}_d$$^{[4]}$} & \boldmath{${\dot m}_h$$^{[4]}$} & \boldmath{$R$$^{[4]}$} & \boldmath{$X_s$$^{[4]}$} &\boldmath{ $M_{BH}$$^{[4]}$} & \boldmath{\textbf{lineE}$^{[5]}$} & \boldmath{\textbf{sigma}$^{[5]}$} & \boldmath{\textbf{norm}$^{[5]}$} &\boldmath{ ${\chi}^2$\textbf{/dof}$^{[6]}$}  \\
& \textbf{(1)}        & \textbf{(2)}      & \textbf{(3)}     & \textbf{(4)}           & \textbf{(5)}           & \textbf{(6)}     & \textbf{(7)}              & \textbf{(8)}            & \textbf{(9)}           & \textbf{(10)} & \textbf{(11)}  & \textbf{(12)} & \textbf{(13)} \\
\midrule
Model 1:      & NU1 & $0.50^{\pm0.12}$ &&& $1.48^{\pm0.03}$ & $0.61^{\pm0.02}$ & $3.46^{\pm0.13}$ & $134^{\pm1}$ & $7.9^{\pm0.03}$ & $6.20^{\pm0.19}$ & $0.80^{\pm0.15}$ & $4^{\pm0.8}$&2121/1496\\
TBabs*(TCAF+  & NU2 & $0.50^{\pm0.12}$ &&& $1.50^{\pm0.08}$ & $0.60^{\pm0.02}$ & $3.65^{\pm0.12}$ & $134^{\pm1}$ &$10.2^{\pm0.08}$ & $6.20^{\pm0.21}$ & $0.79^{\pm0.11}$ & $6^{\pm0.7}$&2079/1497\\ 
Gaussian)     & NU3 & $0.50^{\pm0.13}$ &&& $1.04^{\pm0.04}$ & $0.60^{\pm0.01}$ & $3.42^{\pm0.15}$ & $134^{\pm1}$ & $8.4^{\pm0.30}$ & $6.20^{\pm0.24}$ & $0.70^{\pm0.10}$ & $3^{\pm0.9}$&1767/1435\\
\midrule
Model 2:         & NU1 & $0.50^{\pm0.23}$ & $0.29^{\pm0.03}$ & $0.86^{\pm0.20}$ & $2.10^{\pm.03}$  & $0.67^{\pm0.01}$ & $2.70^{\pm0.08}$ & $132^{\pm1}$ & $7.9^{\pm0.02}$ & $6.30^{\pm0.17}$ & $0.80^{\pm0.24}$ & $5^{\pm0.2}$& 1607/1494\\
TBabs*(reflect*  & NU2 & $0.50^{\pm0.43}$ & $0.32^{\pm0.12}$ & $0.72^{\pm0.14}$ & $2.00^{\pm0.03}$ & $0.63^{\pm0.02}$ & $2.65^{\pm0.11}$ & $125^{\pm3}$ & $7.9^{\pm0.21}$ & $6.31^{\pm0.18}$ & $0.71^{\pm0.24}$ & $3^{\pm0.9}$& 1647/1495\\
TCAF+Gaussian)   & NU3 & $0.83^{\pm0.21}$ & $0.32^{\pm0.05}$ & $0.70^{\pm0.17}$ & $1.50^{\pm0.03}$ & $0.60^{\pm0.02}$ & $2.69^{\pm0.12}$ & $125^{\pm2}$ &$10.1^{\pm0.25}$ & $6.51^{\pm0.14}$ & $0.21^{\pm0.18}$ & $1^{\pm0.1}$& 1694/1433\\
\bottomrule
\end{tabularx}

	\end{adjustwidth}
\noindent{\footnotesize
{$^{[1]}$ ID of the observed dates as described in Table~\ref{tab1} (Col. 1).}
{$^{[2]}$ Model-fitted value of hydrogen column density $n_H$ in $10^{22}$ atoms per cm$^{-2}$ (Col. 2).}
{$^{[3]}$ For \textsc{reflect}*TCAF+Gaussian model, the~value of reflection scaling factor ($rel_{refl}$)}
{and the cosine of inclination angle ($cosIncl$); Col. 3 and Col. 4, respectively.}
{$^{[4]}$ TCAF model-fitted parameters: disk rate (${\dot m}_d$ in Eddington rate ${\dot M}_{Edd}$) in $10^{-3}$ order,} 
{halo rate (${\dot m}_h$ in ${\dot M}_{Edd}$), compression ratio ($R$), shock location ($X_s$ in Schwarzschild radius $r_s$)} 
{and mass of the black hole ($M_{BH}$ in solar mass $M_{\odot}$); Cols. 5--9, respectively.}
{$^{[5]}$ Line energy of the Gaussian line energy (lineE) in~keV, line width (sigma) in~keV} and
{total photons/cm$^2$/s in the line (norm) in $10^{-3}$ order; Cols. 10, 11, and~12, respectively.}
{$^{[6]}$ Model-fitted ${\chi}^2_{red}$; Col. 13 as ${\chi}^2/dof$, where `dof' represents degrees of}
{freedom.}
}
\end{table}

\begin{adjustwidth}{-\extralength}{0cm}
\printendnotes[custom]
\reftitle{References}

\end{adjustwidth}
\end{document}